\documentclass[10pt]{iopart}
\usepackage{iopams}
\usepackage{setstack}

\usepackage[english]{babel} 
\usepackage{amssymb} \usepackage{mathrsfs}
\usepackage{eucal} \usepackage[dvips]{graphicx}
\input xy
\xyoption{all}
\usepackage[vcentermath]{youngtab}
\usepackage{verbatim} \usepackage[usenames]{color}

\newcommand{\defeq}{\doteqdot} 
\newcommand{\reals}{\mathbb{R}} \newcommand{\cplx}{\mathbb{C}}
\newcommand{\proj}{\mathbb{P}}

\newcommand{\ytiny}{\tiny}

\renewcommand{\parallel}{{\rm s}}
\renewcommand{\perp}{{\rm d}}

 \newcommand{\deco}{\mathscr{D}}

\newcommand{\bauth}[1]{#1}
\newcommand{\btitle}[1]{#1} \newcommand{\bjou}[1]{\emph{#1}}
\newcommand{\bvol}[1]{\textbf{#1}} 
\newcommand{\bpage}[1]{#1} \newcommand{\byear}[1]{#1}
\newcommand{\bentry}[6]{\bauth{#1} \byear{#6} \btitle{#2} \bjou{#3}
  \bvol{#4}, \bpage{#5}}

\newcommand{\fpart}{\mathscr{Z}} 

\newcommand{\enel}{\mathscr{F}} 
\newcommand{\DHL}{{\rm{DHL}}}

\renewcommand{\star}{{\textcolor{Gray}{\rule[-0.4pt]{2.4pt}{2.4pt}}}}

\newtheorem{obs}{Note}[section]

\newtheorem{definition}[obs]{Definition}

\newcommand{\dynspace}{\mathscr{M}}
\newcommand{\physspace}{\mathscr{P}}

\newcommand{\hy}{h_{\young(\star)}}
\newcommand{\Hy}{H_{\young(\star)}}
\newcommand{\hn}{h_{\yng(1)}}
\newcommand{\Hn}{H_{\yng(1)}} 
\newcommand{\Jp}{J_\parallel}
\newcommand{\Jnp}{J_\perp}

\begin{document} 
\title[Potts models on HL and RG dynamics II: examples and numerical results]{Potts models on hierarchical lattices and Renormalization Group dynamics II: examples and numerical results}
\author{Jacopo De Simoi}
\address{University of Maryland, College Park, MD 20740, USA}
\ead{jacopods@math.umd.edu}

\begin{abstract}
We obtain the exact renormalization map and plots of Lee-Yang and Fisher zeros distributions for Potts models on a number of hierarchical lattices: the diamond hierarchical lattice, a lattice we call spider web, the Sierpinski gasket and cylinders. Such models are only examples among the ones we can study in the general framework of hierarchical lattices, developed in a previous paper.  
\end{abstract}
\section{Introduction} 
Spin models on hierarchical lattices are a large class of exactly soluble models that have been first considered as approximations to regular lattices \cite{Mi1,Mi2,Kad} and then as examples of lattices invariant under a real-space renormalization procedure \cite{boh,kg1,kg3,kg2}. The renormalization group action for such models is therefore exact and the study of its dynamics provides some interesting results that can be useful in studying the renormalization group action in more general cases. In this paper we consider some examples of Potts models on hierarchical lattices, namely the diamond hierarchical lattice (\sref{dhl}), the spider web (\sref{spider}), the Sierpinski gasket (\sref{sierpinski}) and cylinders (\sref{pipes}). Apart from the case of the spider web which (to the best of the author's knowledge) has not yet been subject of research, models on the other lattices have been extensively studied before (e.g. \cite{dsi,zal,bly},\cite{gasm,bcd}); however, the purpose of this paper is to present all such models with a consistent and uniform method which also allows for the presence of an external magnetic field. This approach has been presented in a previous paper \cite{pap1} and can be applied in full generality to all hierarchical lattices. For each model we will write the exact renormalization group generator and provide numerical results for the distribution of Lee-Yang and Fisher zeros. Such results are obtained using techniques which we explain in \sref{numerical} and in the appendices. We also report some observations which arise quite naturally from the analysis of the aforementioned models and also provide some new results. In particular, we observe that Lee-Yang zeros responsible for the infinite susceptibility of Ising model on the diamond hierarchical lattice in the paramagnetic phase are given by interactions that are only finitely renormalizable. Moreover, we are able to write the exact renormalization map associated to the Potts model on a Sierpinski gasket for all values of $q$. We refer the interested reader to \cite{pap1} for a detailed treatment of Potts models on hierarchical lattices; what follows in this introduction is an attempt to summarize consistently all basic concepts we need in this paper.\\
 Hierarchical lattices are infinite lattices in which we allow multiple-spin connections and that are obtained by iterating a \emph{decoration} procedure on a finite lattice; this procedure amounts to substitute each edge of a lattice with a given block of spins and edges (see e.g. figures \ref{fig:dhlb},\ref{fig:spider},\ref{figure:sierpinski}). In \cite{pap1} we showed that, for such models, we can define a renormalization map that acts as a \emph{polynomial map} on the complex vector space of Boltzmann weights $\exp(-\beta J)$ associated to local interactions $J$. We observe that a different normalization of the Boltzmann weights or, equivalently, a different choice of zero of energies, does not change the thermodynamics of the models. Therefore, we argue that the space of Boltzmann weights can be considered as a projective space $\proj^r$ and the renormalization map will act on such projective space as a \emph{rational map}. In general, if we consider models with several type of interactions, then the renormalization map will act on the Cartesian product of several projective spaces (i.e a so-called \emph{multiprojective space}) that we call \emph{dynamical space}. We define \emph{physical space} to be the space of Boltzmann weights associated to interactions given by pair interactions and (possibly) by coupling with an external magnetic field. In general, the physical space is a submanifold of the dynamical space which is \emph{not} preserved by the renormalization map. This amounts to the well-known fact that the renormalization of pair interactions introduces new multiple-spin interactions. Hierarchical lattices are such that all possible multiple-spin interactions that arise from the renormalization process form a finite-dimensional space; in this sense we say that hierarchical lattices are exactly renormalizable.
\section{Numerical approaches}\label{numerical} 
\Yboxdim{2.75pt}\Ylinethick{0.25pt}
In the next sections we will perform a numerical study of rational maps that generate the renormalization group of some examples of hierarchical lattices. We are, in fact, interested in finding the distribution of Lee-Yang and Fisher zeros for such models. Given the renormalization map of the model, one method to obtain numerically such distributions is to find all basins of attraction of stable fixed points of the map; the boundary of such regions is going to be the unstable set for the renormalization map (i.e. the so-called Julia set) and phase transitions of the model will appear for interactions belonging to such set (see \ref{basins} for more details). A second approach, in some sense more straightforward, proceeds by computing an approximation of a real function called \emph{Green function}. This function is a purely dynamical object and it is related to the \emph{free energy} of the model the map is associated to; in particular we expect the two functions to have the same domain of analyticity (although this fact has been formally proved only for some cases). Once we obtain the numerical approximation to the Green function, applying the Laplacean differential operator yields the density of the measure supported on the Lee-Yang and Fisher zeros of the model (see \ref{green} for details).\\
As we pointed out in the introduction, the renormalization group action on Boltzmann weights is generated by a rational map on a multiprojective space $\dynspace$ called \emph{dynamical space} which contains all multiple-spin interactions that can be generated by the renormalization process. We will often consider a submanifold $\physspace$ that we call \emph{physical space}. This submanifold is given by Boltzmann weights associated to interaction that are induced by pair interactions and possibly an external magnetic field. 
Let us define the pair interactions, i.e. let $\Jp$ be the energy given to two parallel neighbouring spins and $\Jnp$ the energy associated to two neighbouring spins that are in different states. The Boltzmann weights associated to the corresponding energies will be denoted by $[z_\parallel:z_\perp]$ and belong to the one-dimensional complex projective space $\proj^1$. A magnetic field, if present, will assign energy $H_{\young(\star)}$ to one \emph{special} state among the $q$ Potts states and energy $H_{\young(\hfil)}$ to all other states. The Boltzmann weights associated to the corresponding energies will be denoted by $[h_{\young(\star)}:h_{\young(\hfil)}]\in\proj^1$.\\
For each hierarchical lattice we can therefore define a map from $\proj^1\times\proj^1\to\physspace\subset\dynspace$ that gives projective coordinates to $\physspace$. The pair interaction Boltzmann weights will belong to the first $\proj^1$ factor and the magnetic field weight to the second $\proj^1$ factor. 
We are now going to define standard local charts (coordinates) on each $\proj^1$ factor; all numerical computations will be performed in one of such charts. Notice that all the coordinates we are going to define are just standard (inhomogeneous) charts on the projective line $\proj^1$. 
\begin{definition}
We call \emph{standard interaction coordinates} the coordinate chart of $\proj^1$ given by $\zeta\defeq z_\parallel/z_\perp=\exp\left(-\beta\left(\Jp-\Jnp\right)\right)$ (for $z_\perp\not = 0$).\\
When dealing with zero-temperature phase transitions we will need to consider the inverse chart, $1/\zeta$ (for $z_\parallel\not = 0$); we will call such chart \emph{inverse interaction coordinates}.
\end{definition}
Note that the standard interaction coordinates could be obtained by setting $\Jnp=0$ and considering the Boltzmann weights $\zeta$ corresponding to such choice of zero of energies. In this sense we call them \emph{standard}. In such coordinates, $\zeta=0$ and $\zeta=\infty$ are respectively the antiferromagnetic and ferromagnetic points, while $\zeta=1$ is the paramagnetic point. The latter is fixed by all renormalization maps, while the ferromagnetic point ($\zeta=\infty$) is fixed whenever the hierarchical lattice is connected; the antiferromagnetic point is usually mapped to the ferromagnetic point by the RG map.
\Yboxdim{3.0pt}\Ylinethick{0.25pt}
\begin{definition}
We call \emph{standard field coordinates} the coordinate $h\defeq\hy/\hn=\exp\left(-\beta\left(\Hy-\Hn\right)\right)$.
\end{definition}
Again, the standard field coordinates can be obtained as the Boltzmann weights associated to the choice $\Hn=0$. In the standard field coordinates, $h=1$ corresponds to the case with zero field, $h=\infty$ corresponds to the case of infinite field and $h=0$ is when the privileged state costs infinite energy and it is therefore never assumed.
\section{Diamond hierarchical lattices}\label{dhl}
\begin{figure}[!htbp]
  \includegraphics[height=2.5cm]{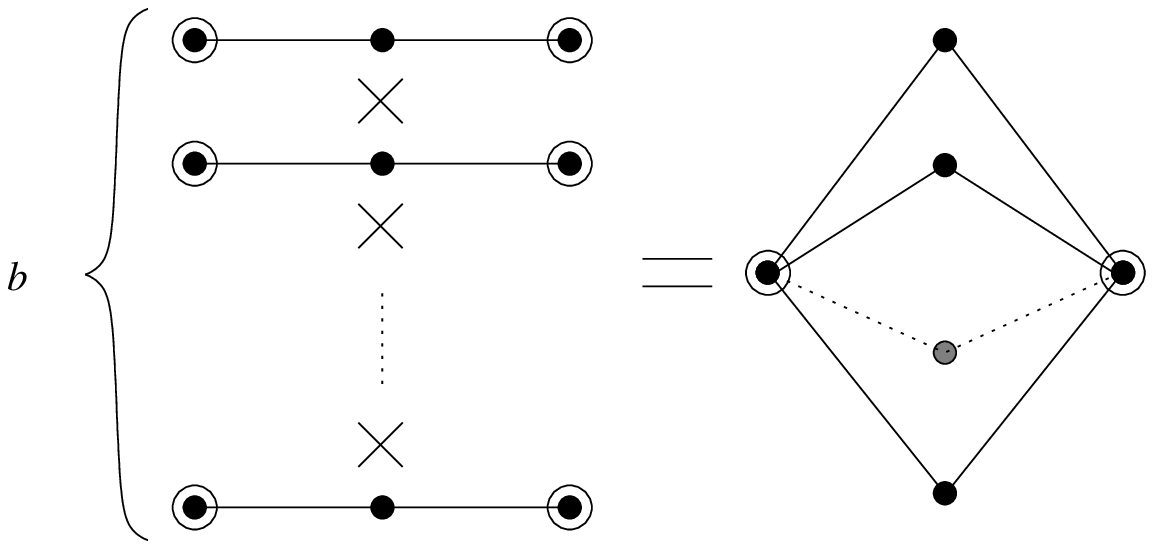}\hspace{1cm}   
   \includegraphics[height=2.5cm]{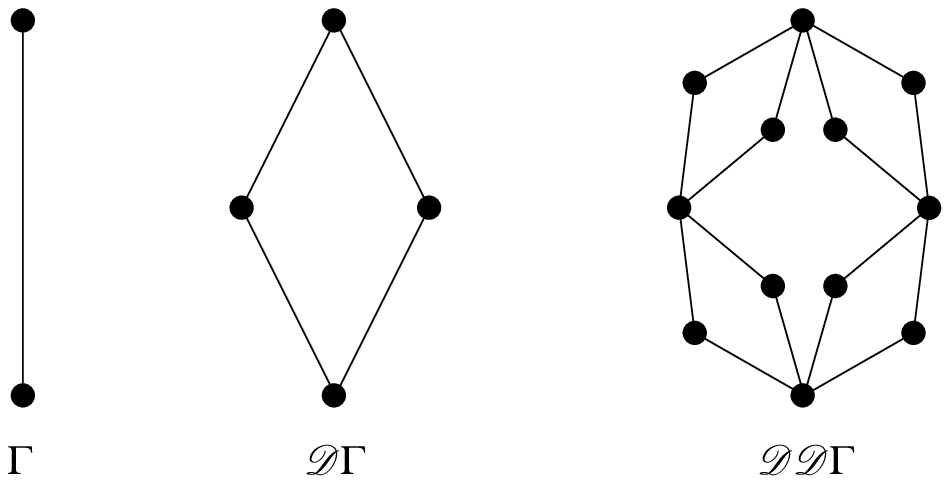}
   \caption{Decoration generating $\DHL_b$ (left) with some iterations of the decoration procedure for $\DHL_2$ on a starting graph $\Gamma$ (right). The hierarchical lattice $\DHL_b$ is the limit graph that we obtain by indefinitely iterating the decoration procedure.}
   \label{fig:dhlb}
 \end{figure}

\begin{figure}[!t]
  \centering
  \includegraphics[width=3.5cm]{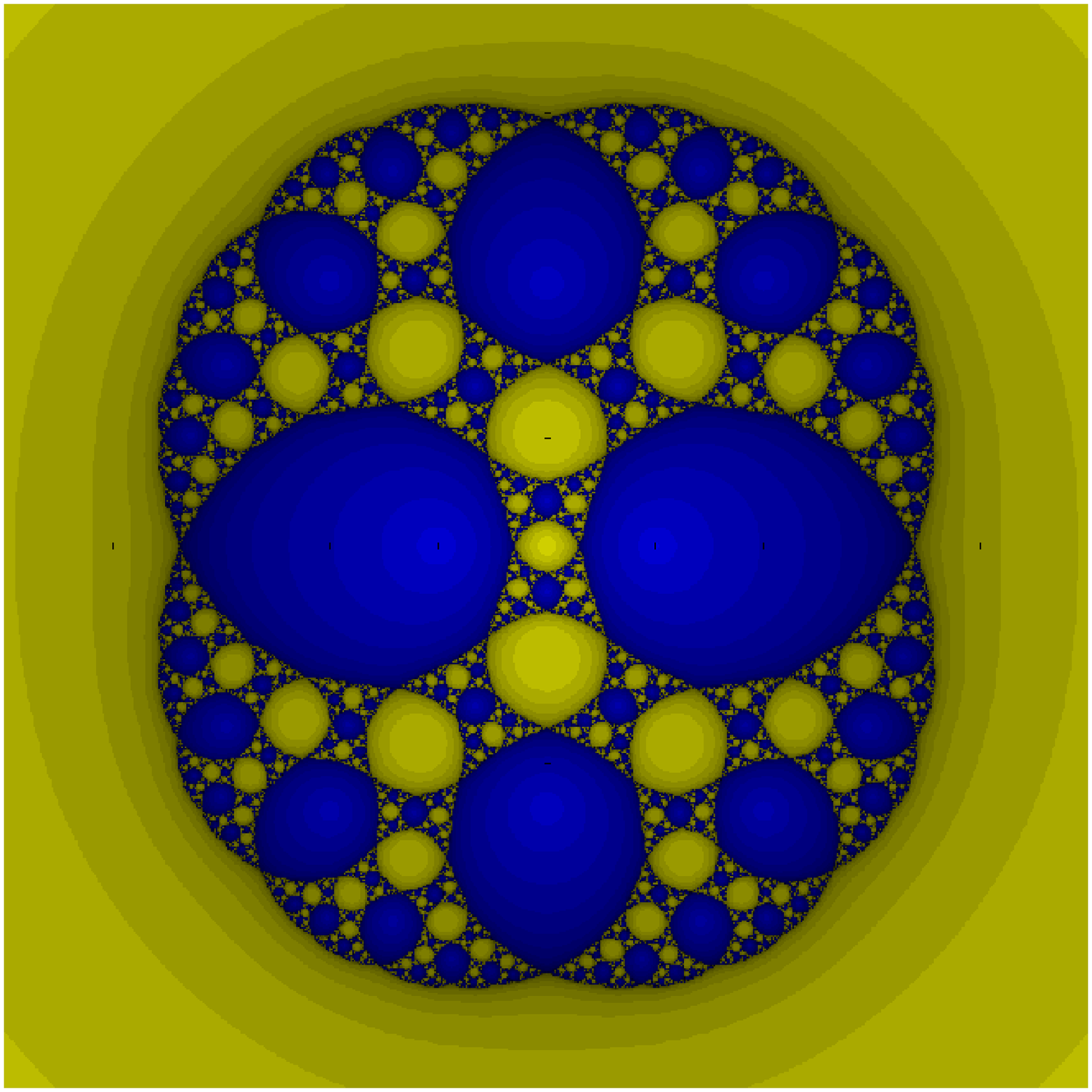}
  \includegraphics[width=3.5cm]{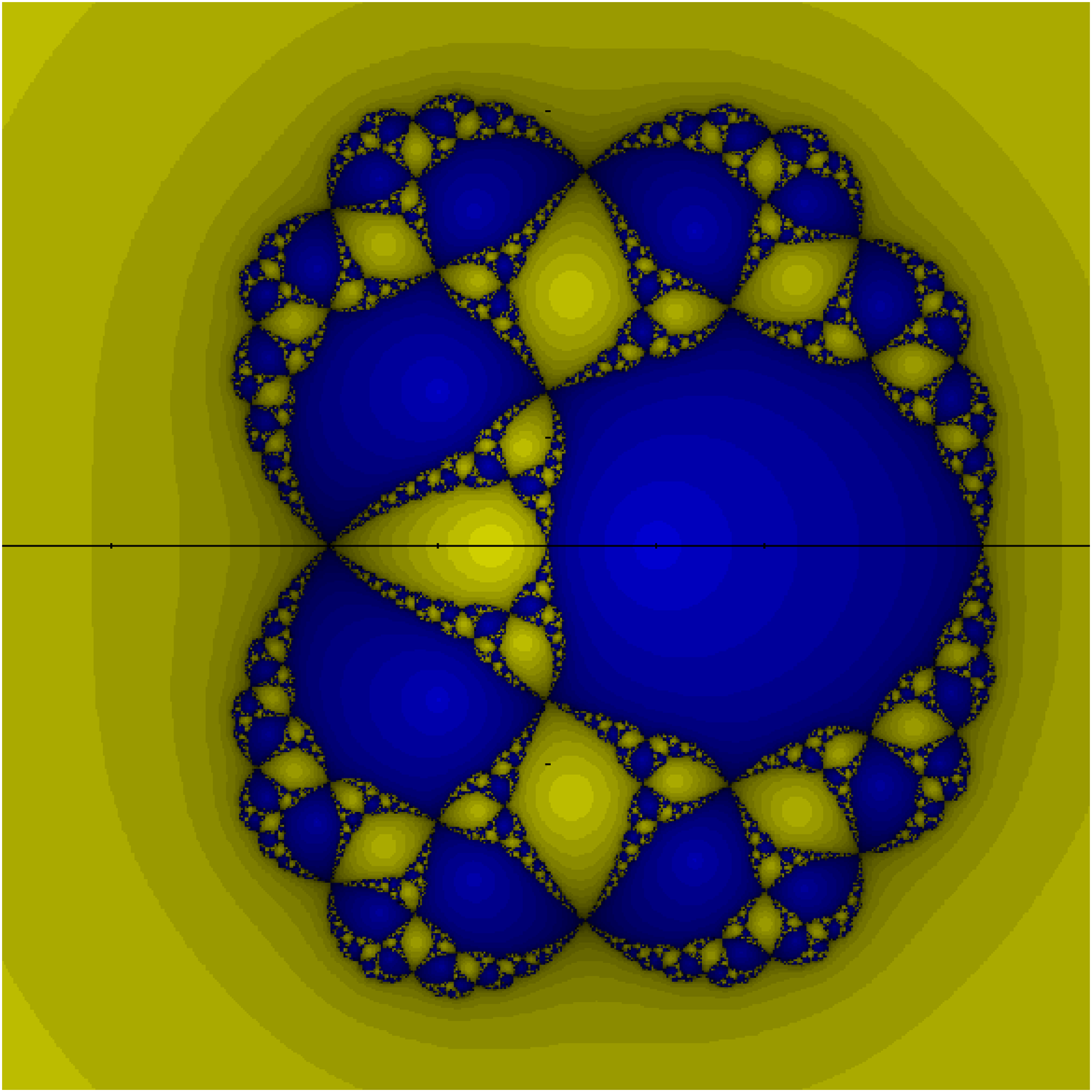}
  \includegraphics[width=3.5cm]{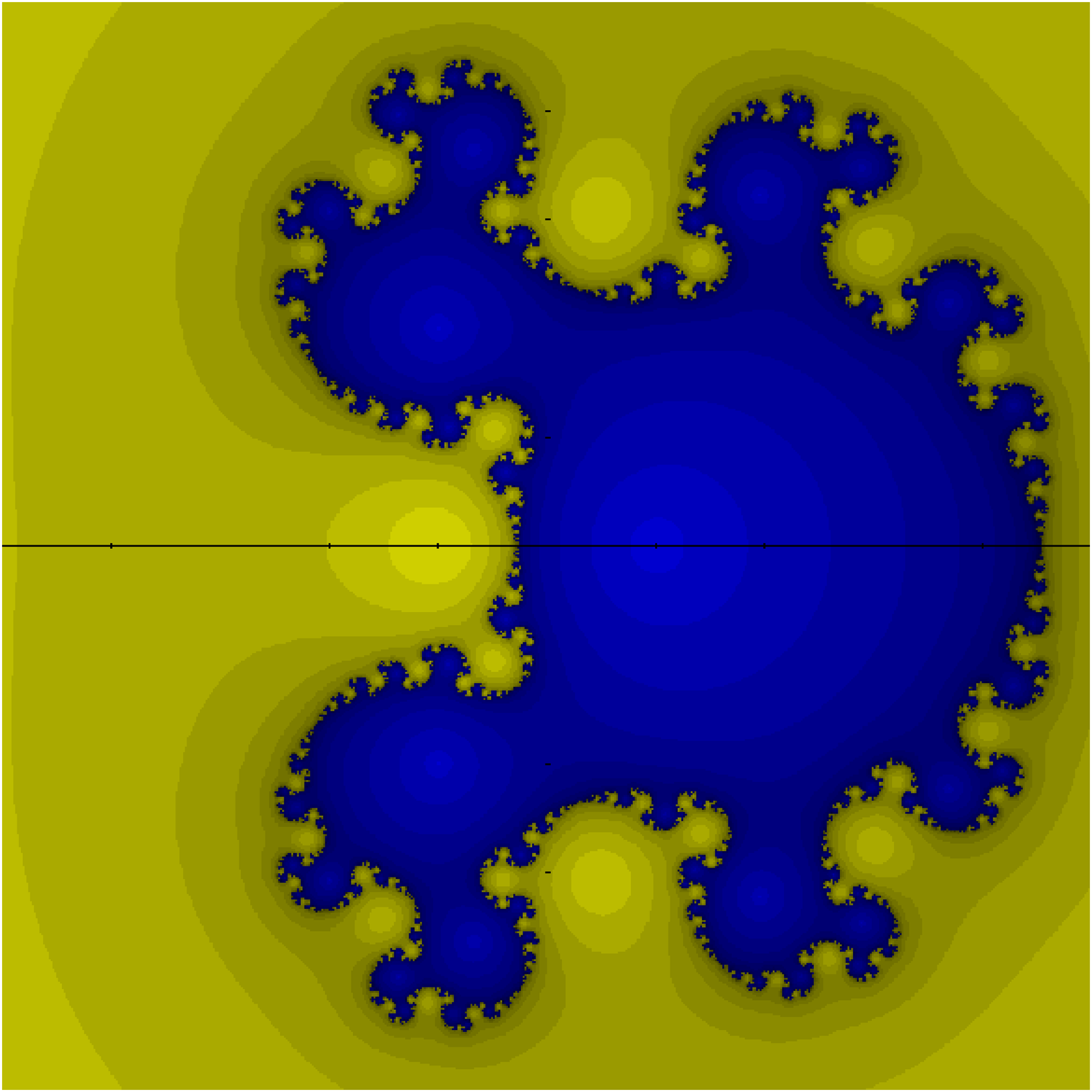}\\
  \includegraphics[width=3.5cm]{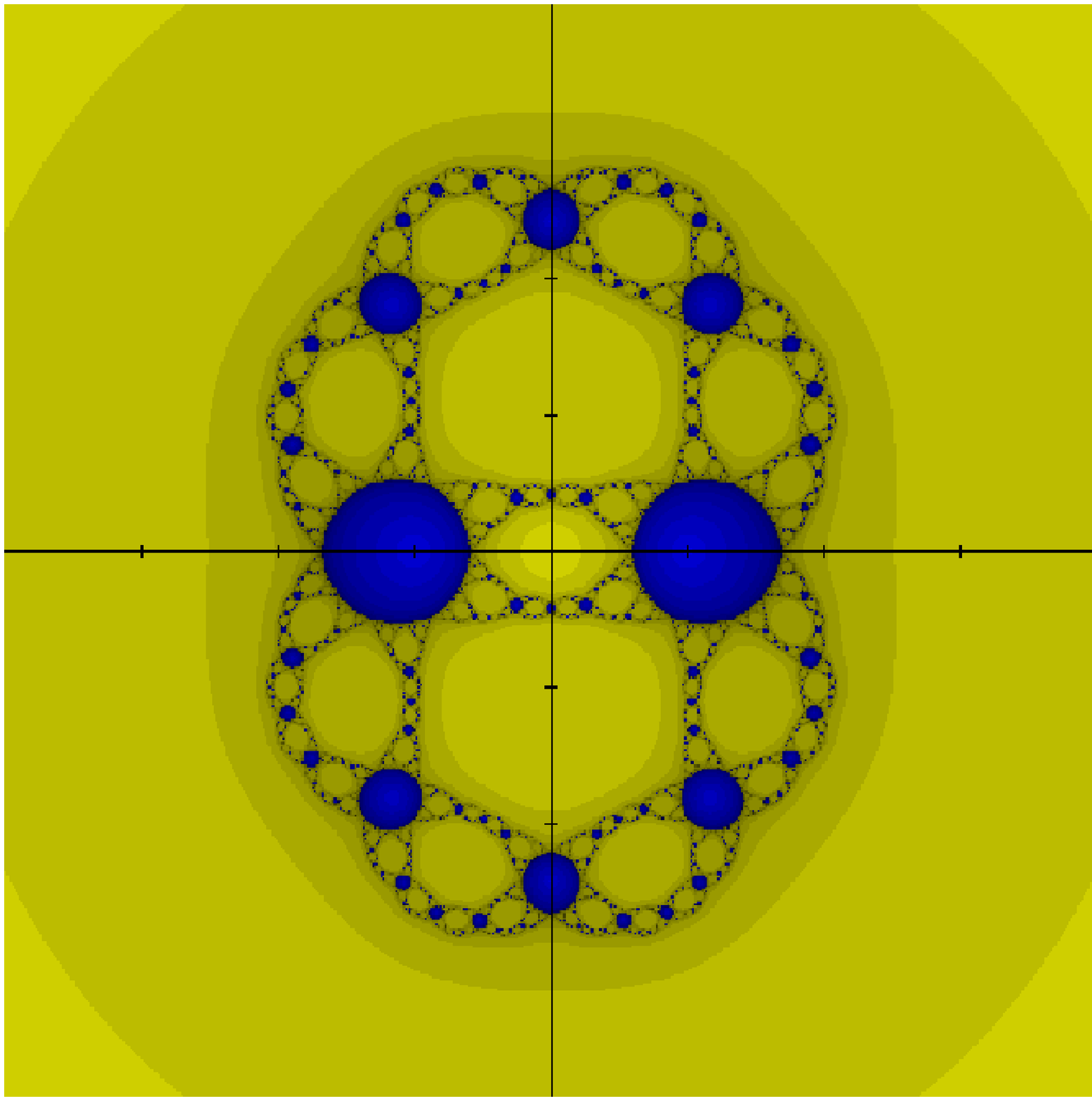}
  \includegraphics[width=3.5cm]{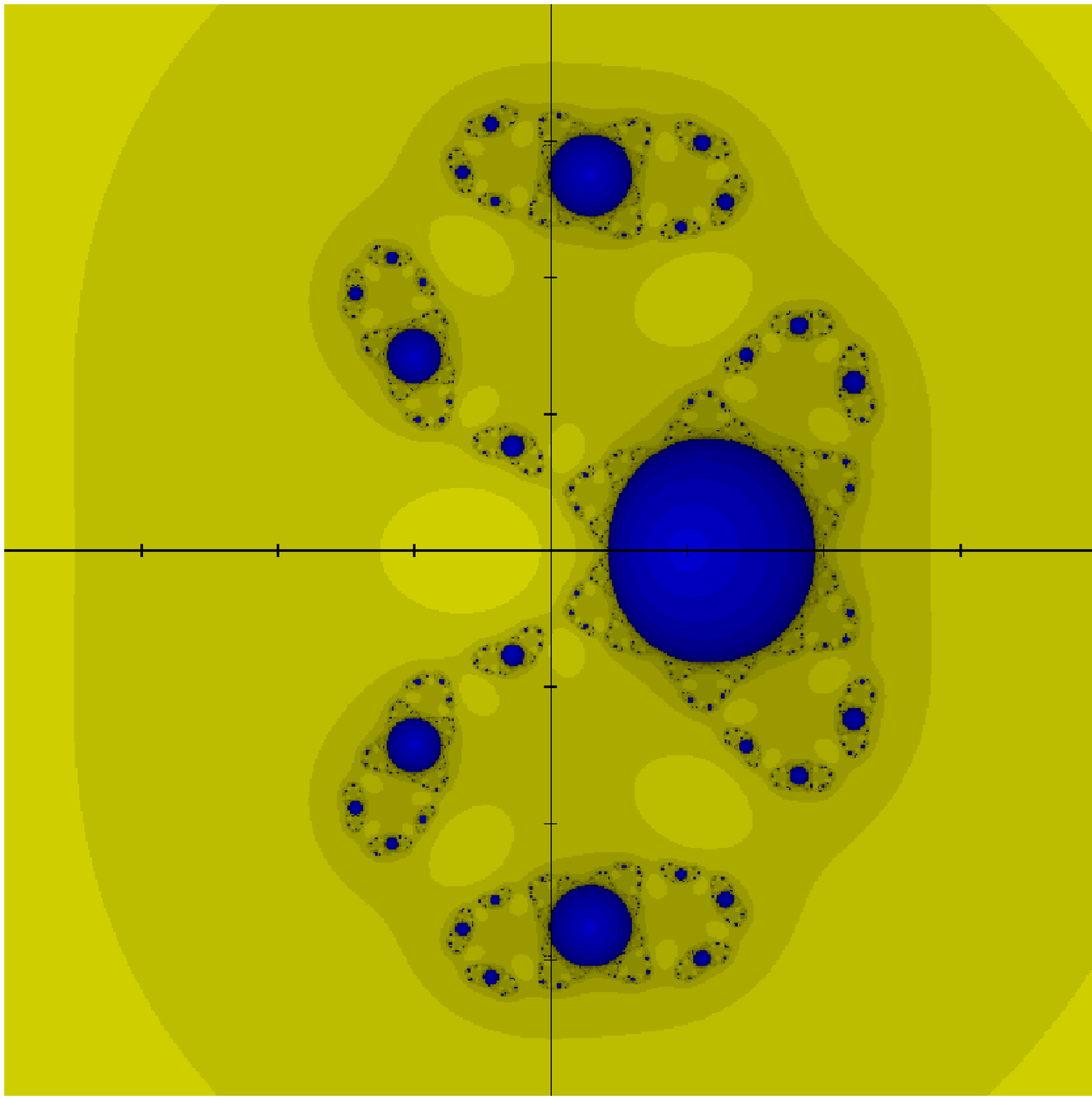}
  \includegraphics[width=3.5cm]{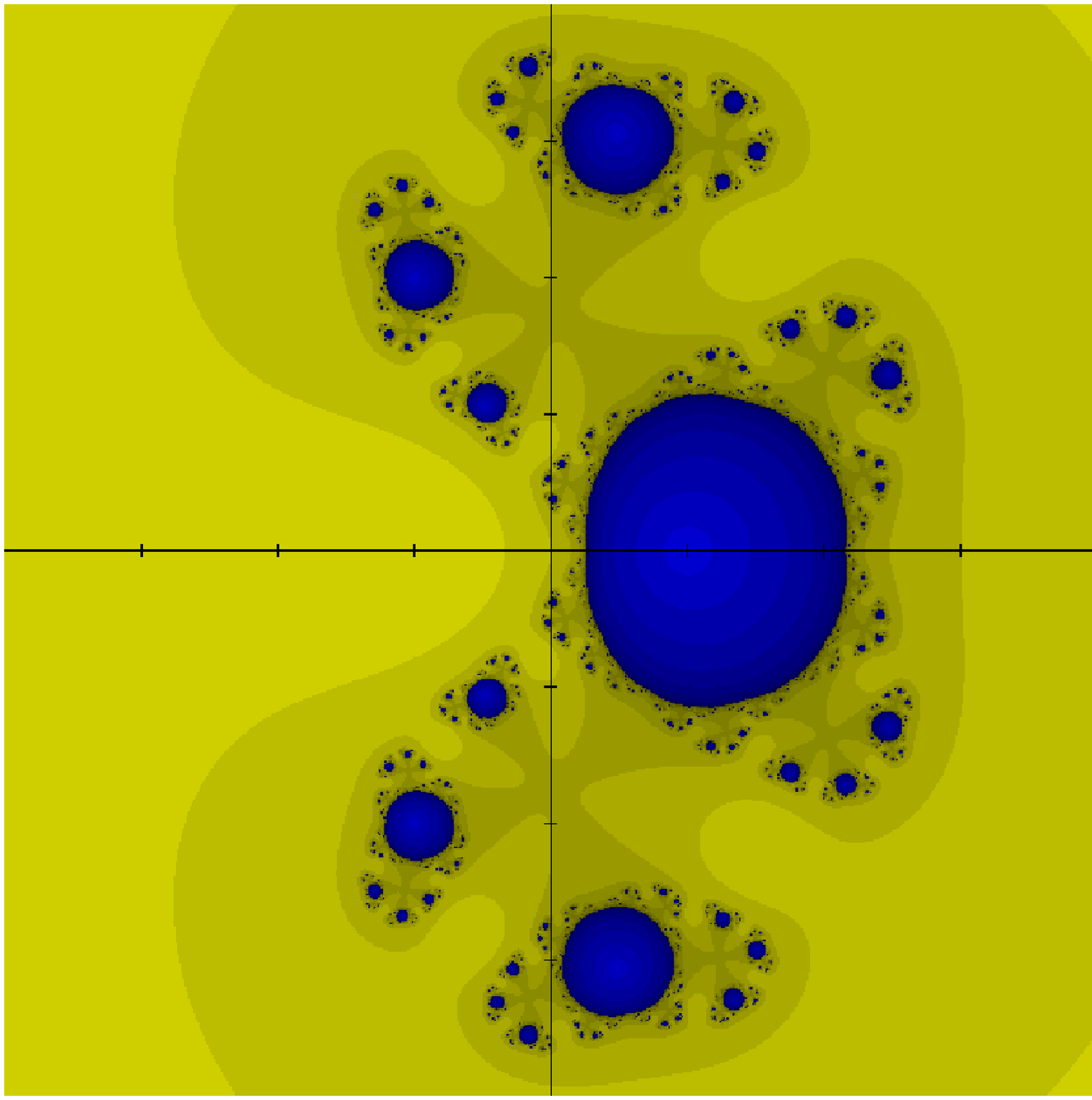}\\
  \includegraphics[width=3.5cm]{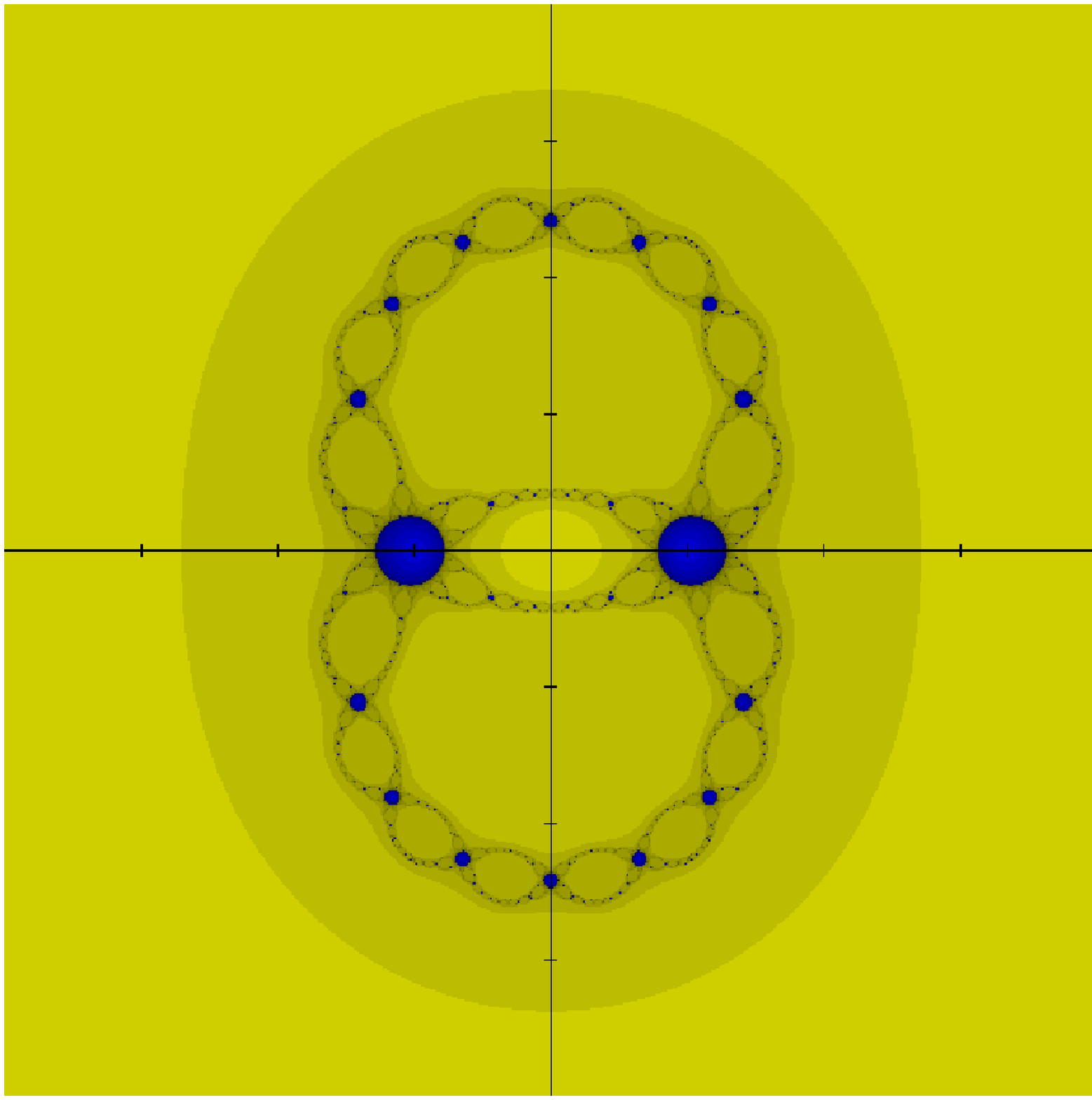}
  \includegraphics[width=3.5cm]{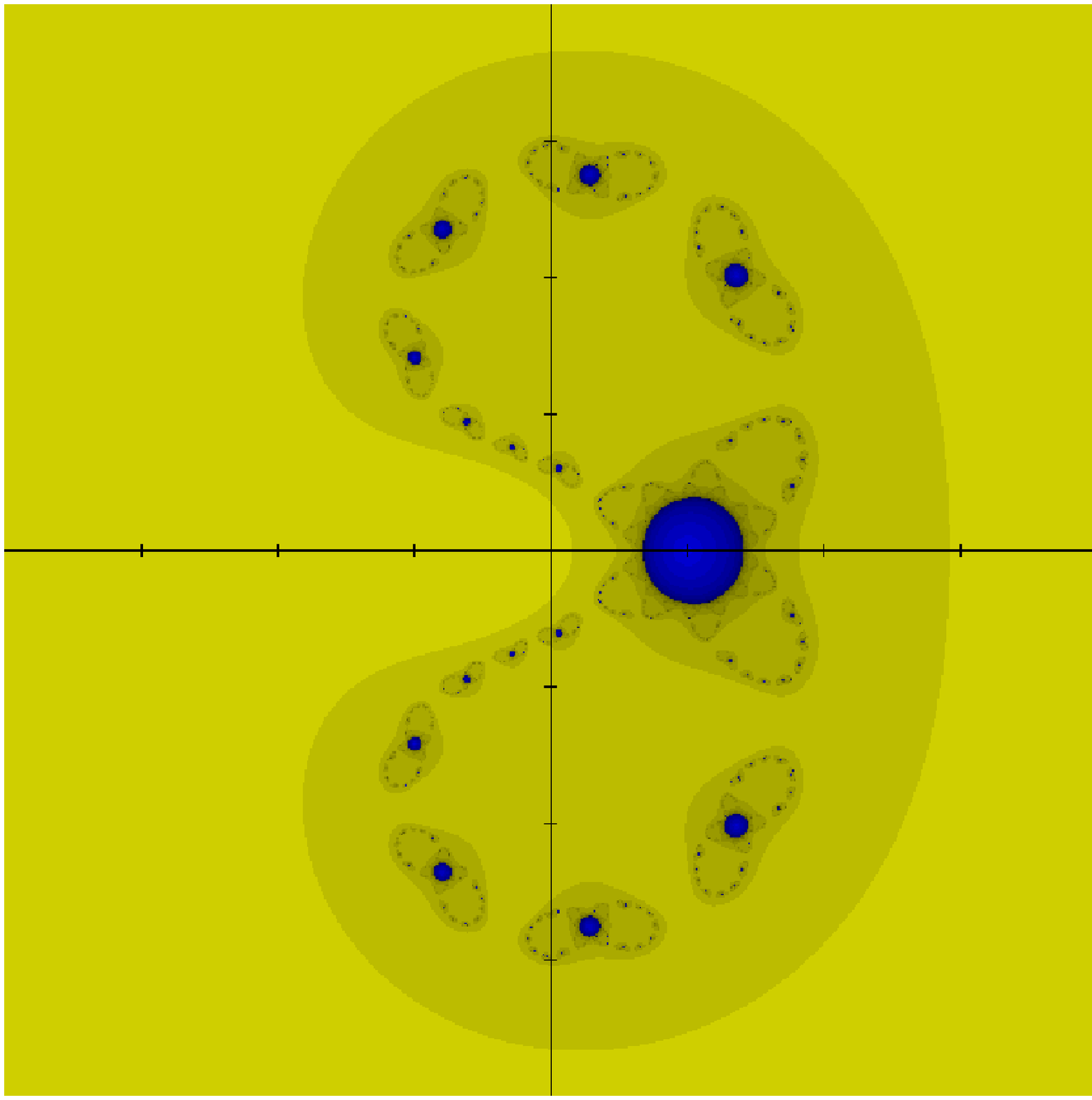}
  \includegraphics[width=3.5cm]{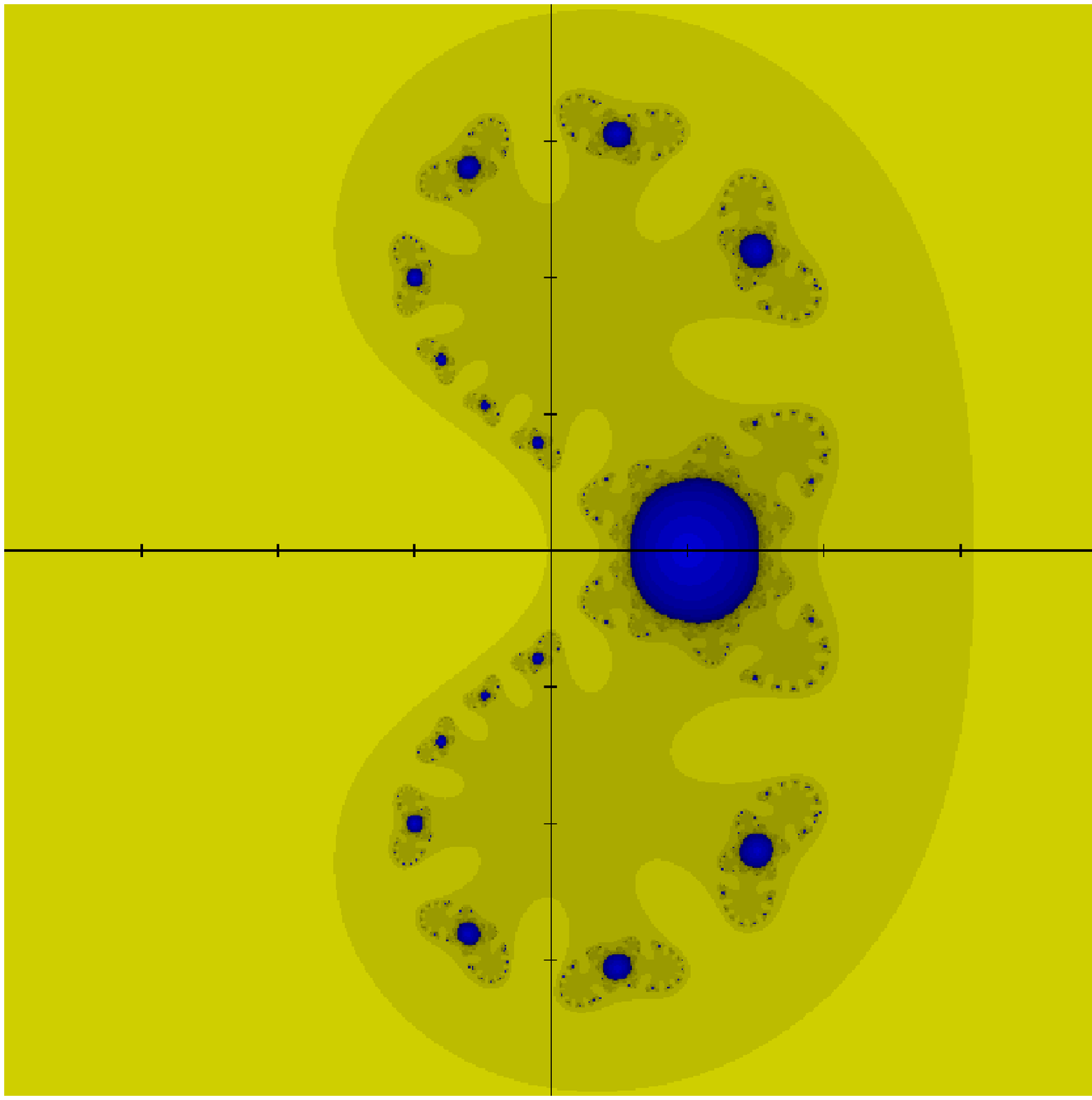}\\
  \includegraphics[width=3.5cm]{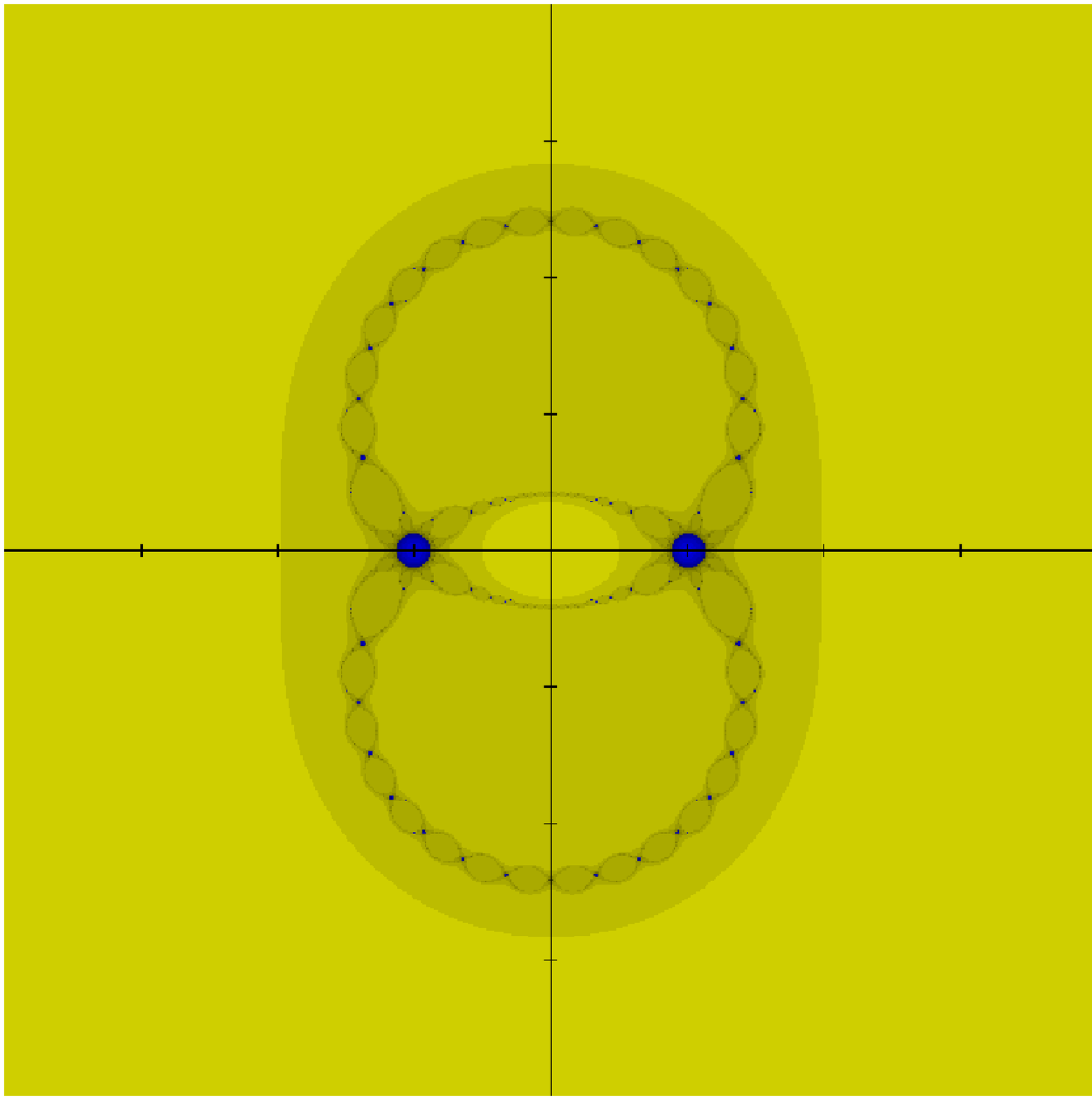}
  \includegraphics[width=3.5cm]{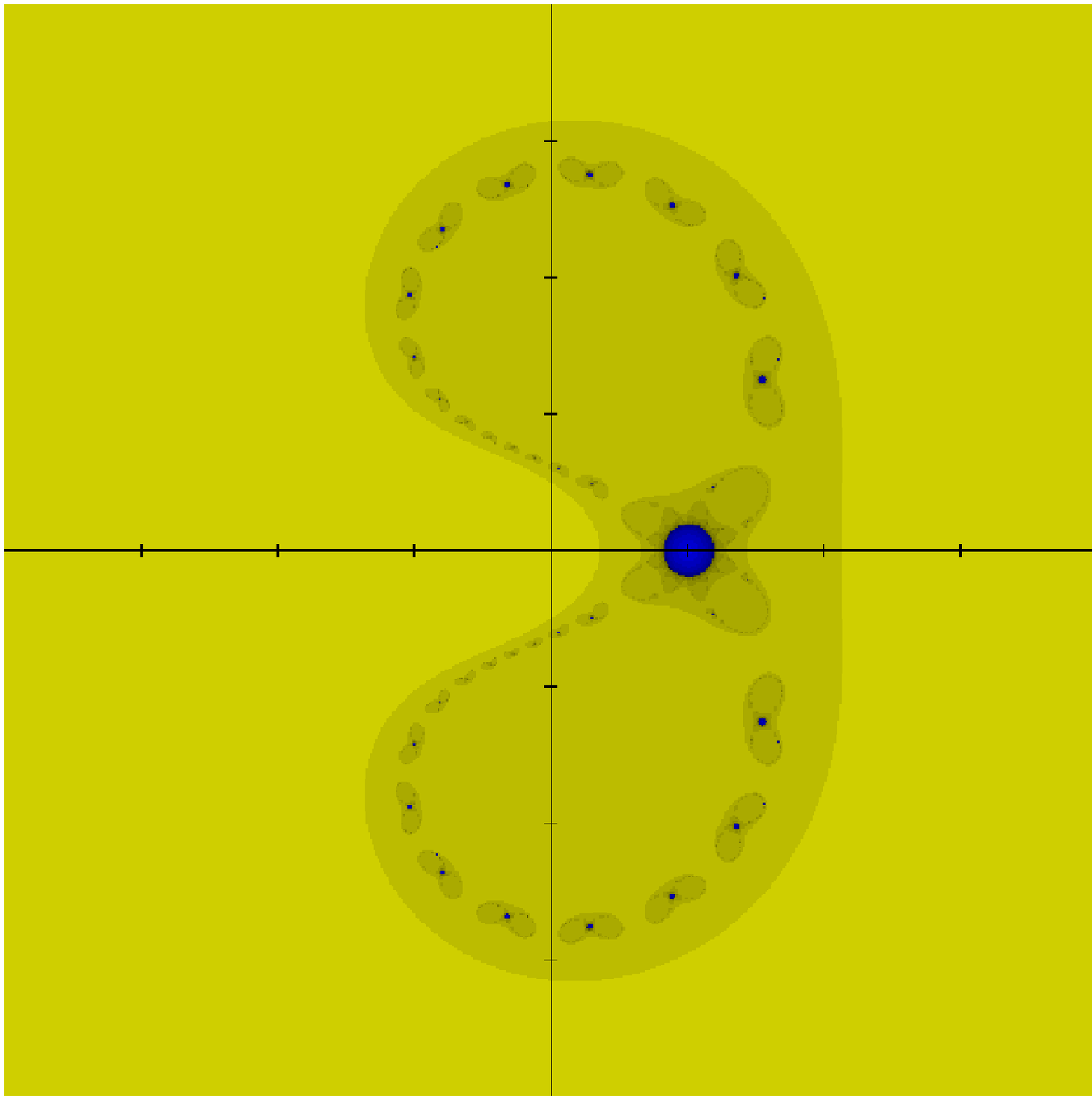}
  \includegraphics[width=3.5cm]{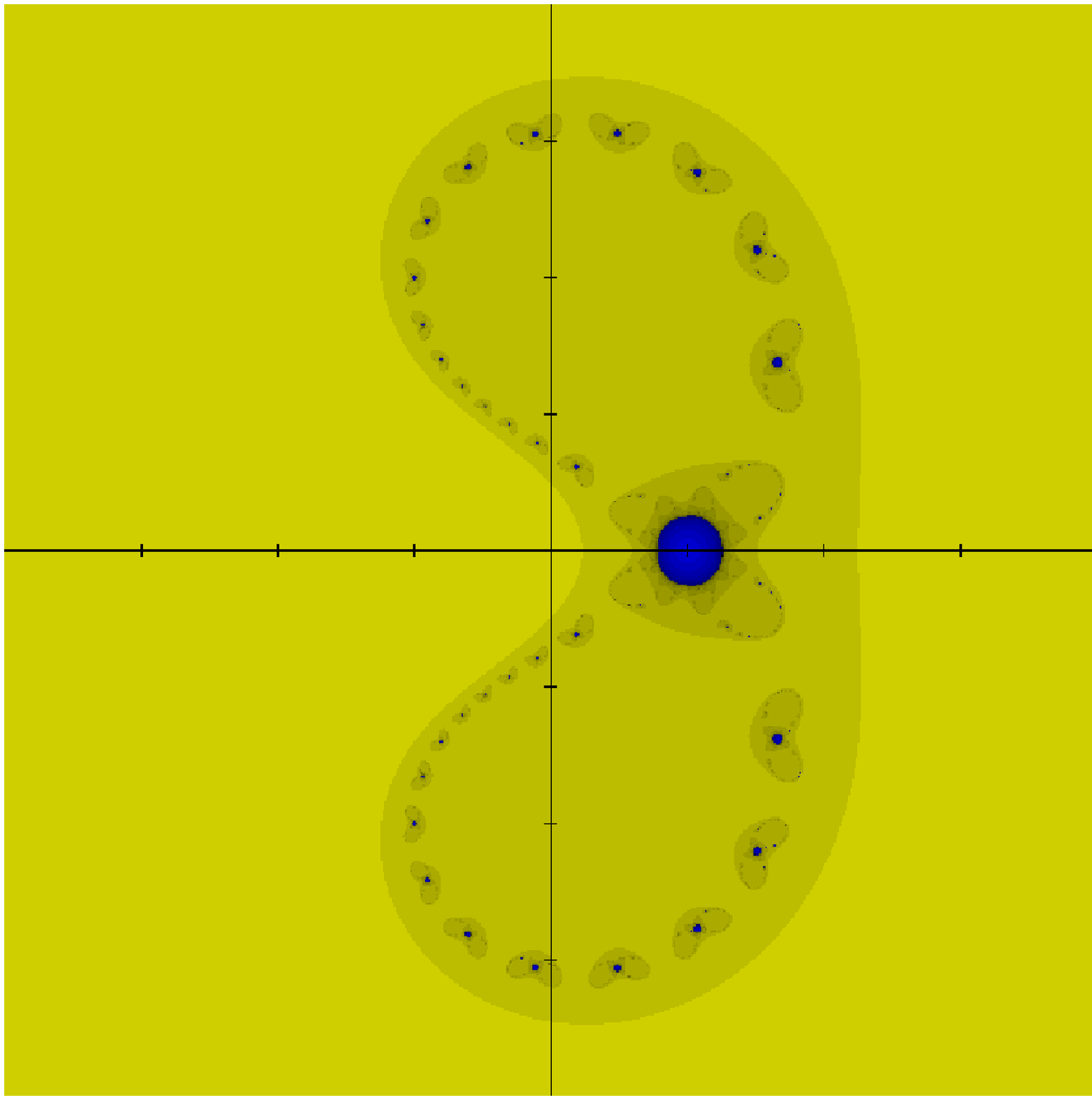}\\
\caption{\emph{Standard interaction coordinates}: Basins of attraction of the paramagnetic fixed point (in blue) and the ferromagnetic one (yellow) for the $\DHL$ for various values of $b$ and $q$; rows have respectively $b=2,4,8,16$, and columns have $q=2,3,4$}
\label{dhlvarib}
\end{figure}

Diamond hierarchical lattices (DHLs) have been the first hierarchical lattices to be investigated using tools from complex dynamics \cite{dsi}; they are lattices on a standard graph, and they can be obtained by iterating the \emph{decoration procedure} illustrated in figure \ref{fig:dhlb}.
\Yboxdim{2.75pt}\Ylinethick{0.25pt}
Recall that we define the interaction as $\Jp$ if two neighbouring spins are in the same state and $\Jnp$ if they are in different states. For this lattice the dynamical variables with no magnetic field are the Boltzmann weights $[z_{\yng(2)}:z_{\yng(1,1)}]=[z_{\parallel}:z_\perp]\in\proj^1$ relative to the states of two neighbouring spins:
\begin{eqnarray*}
z_{\yng(2)}&=\exp(-\beta \Jp) &\qquad{\rm same\ state }\\
z_{\yng(1,1)}&=\exp(-\beta \Jnp) &\qquad {\rm  different\ states}
\end{eqnarray*}  
In this case the physical space of the model without external field coincides with the dynamical space. The renormalization map can be easily written in the dynamical variables for all values of $q$; let $\fpart_{\yng(2)}$ and $\fpart_{\yng(1,1)}$ are the renormalized variables; then:
\numparts
\begin{eqnarray}
\fpart_{\yng(2)}&=&\left(z_{\yng(2)}^2+(q-1)\cdot z_{\yng(1,1)}^2\right)^b\\
\fpart_{\yng(1,1)}&=&\left(2\cdot z_{\yng(2)}\cdot z_{\yng(1,1)}+(q-2)\cdot z_{\yng(1,1)}^2\right)^b
\end{eqnarray} 
\label{dhlnof} 
\endnumparts
For every $b$ the map has a fixed point at $[1:1]$ (paramagnetic point) and at $[1:0]$ (ferromagnetic point). Another fixed point appears when $b$ is odd at $[-1:1]$. The fixed point $[1:1]$ is always superattracting (i.e. the map has zero derivative at the fixed point), while the ferromagnetic fixed point is superattracting only if $b>1$, therefore excluding the one-dimensional chain case. Thus, in all other cases, and for all values of $q$, we expect a phase transition at \emph{finite} temperature, since basins of attraction of an attracting fixed point of a rational map always contain a neighbourhood of the fixed point.
In figure \ref{dhlvarib} we show the aforementioned basins of attraction of the ferromagnetic and paramagnetic fixed point for various values of $b$ and $q$.
\subsection{Magnetic field}
\begin{figure}[!h]
\centering
  \includegraphics[width=4.5cm]{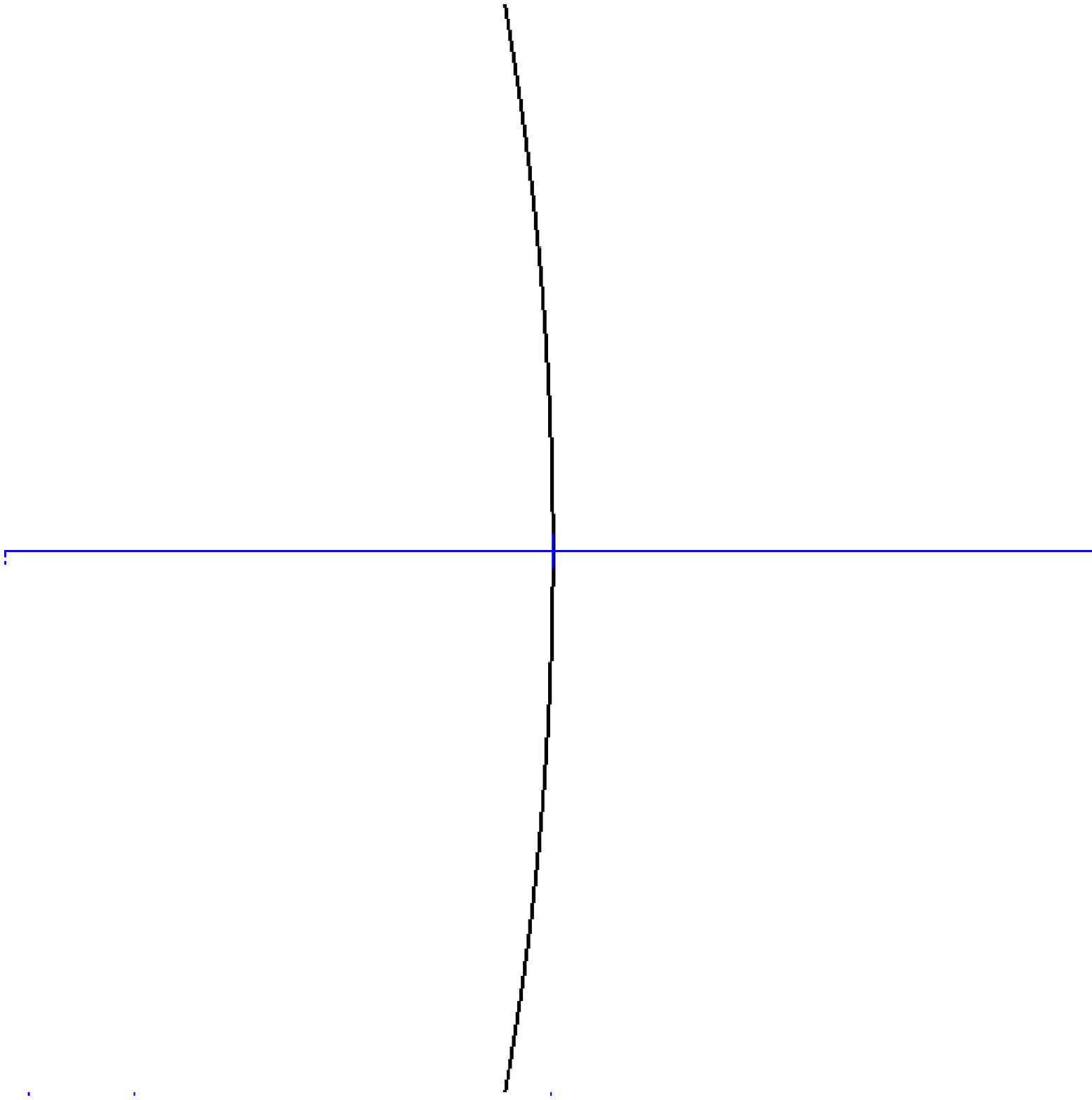}\hspace{1cm}
  \includegraphics[width=4.5cm]{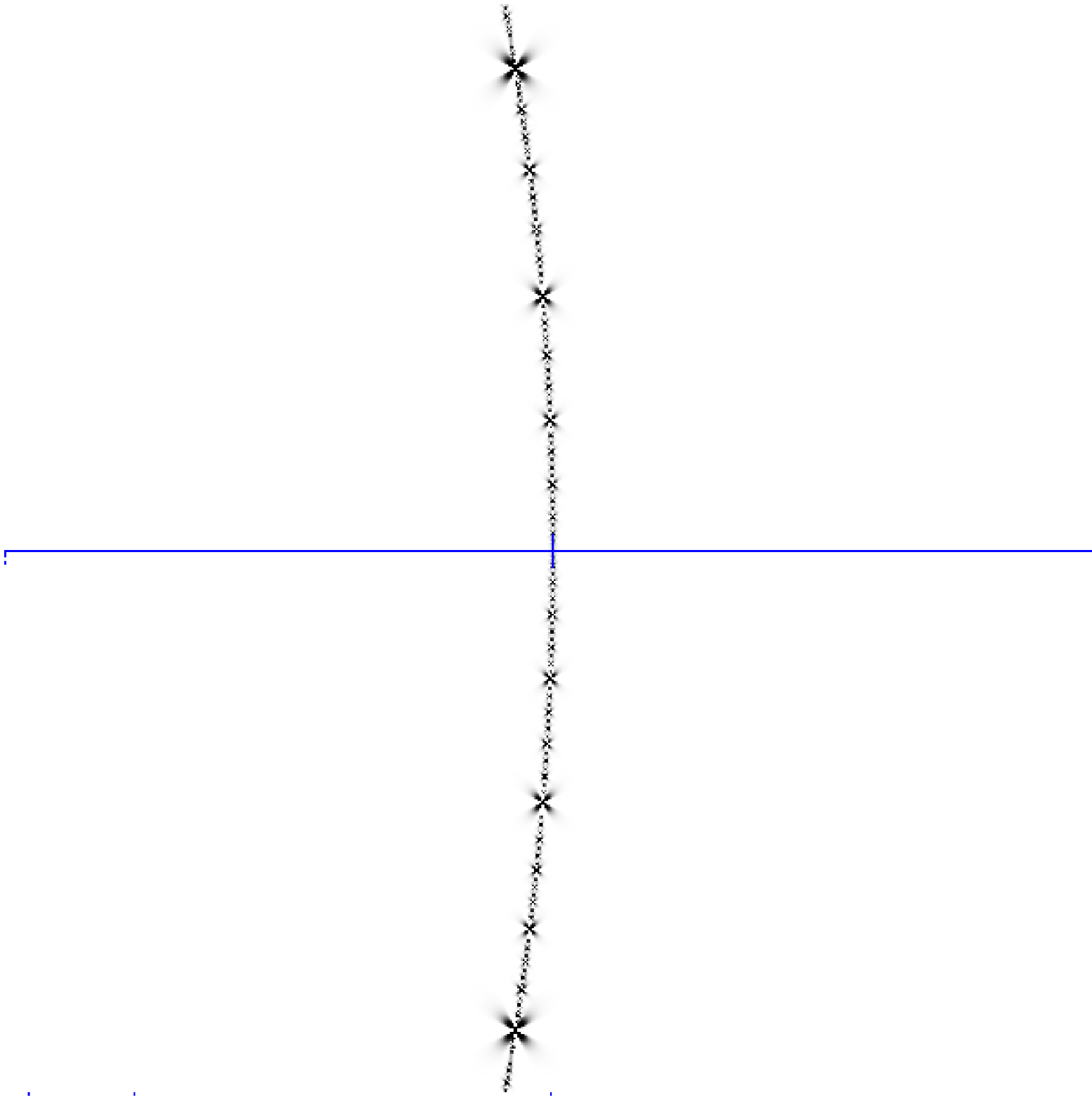}
\caption{Standard field coordinates: Lee-Yang zeros of the diamond hierarchical lattice with $b=2$; the center of both figures is z. Left: zeros in the real ferromagnetic phase; right: the anomalous zeros for the supposedly paramagnetic phase}
\label{criticalphase}
\end{figure}

\begin{figure}[!h]
\centering
\includegraphics[height=5cm,angle=270]{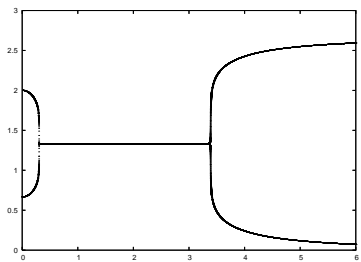}
\caption{Spontaneous magnetization for the Diamond Hierarchical Model ($b=2$). The horizontal axis corresponds to real values of $z$, on the $y$ axis we have spontaneous magnetization (in arbitrary units). We notice the presence of the three phases: antiferromagnetic, critical (paramagnetic) and ferromagnetic.}
\label{magnetization} 
\end{figure}

As explained in the previous paper it is possible to deal in a completely analogous way with an applied magnetic field; corresponding Boltzmann weights  will appear as parameters of the renormalization map. In this case the dynamical variables are the Boltzmann weights $[z_{\young(\star\star)}:z_{\young(\star,\hfil)}:z_{\young(\hfil\hfil)}:z_{\young(\hfil,\hfil)}]\in\proj^3$ relative to the states of two neighbouring spins according to the following rules:
\begin{eqnarray*}
z_{\young(\star\star)} &\qquad{\rm same\ state\ (special)}\\
z_{\young(\star,\hfil)}&\qquad{\rm different\ states (one\ special)}\\
z_{\young(\hfil\hfil)}&\qquad{\rm same\ state(not\ special)}\\
z_{\young(\hfil,\hfil)}&\qquad{\rm different\ states(not\ special)}\\
\end{eqnarray*}    
Now given $\Jp,\Jnp,\Hy,\Hn$ we can define the physical space as given by:
\begin{equation}\eqalign{
z_{\young(\star\star)}=\exp(-\beta(\Jp+2\Hy))\qquad z_{\young(\star,\hfil)}=\exp(-\beta(\Jnp+\Hy+\Hn)) \cr
z_{\young(\hfil\hfil)}=\exp(-\beta(\Jp+2\Hn))\qquad z_{\young(\hfil,\hfil)}=\exp(-\beta(\Jnp+2\Hn))}.
\label{initialDHL}
\end{equation}
The renormalization map is given in \eref{dhlfield1}-\eref{dhlfield2}: 
\numparts
\begin{eqnarray}
\label{dhlfield1} 
\fpart_{\young(\star\star)}&=&\left(h_{\young(\star)}\cdot z_{\young(\star\star)}^2+(q-1)\cdot h_{\yng(1)}\cdot z_{\small\young(\star,\hfil)}^2\right)^b\\
\fpart_{\young(\star,\hfil)}&=&\left(h_{\young(\star)}\cdot z_{\young(\star\star)}\cdot z_{\young(\star,\hfil)}+h_{\yng(1)}\cdot z_{\young(\star,\hfil)}\cdot z_{\young(\hfil\hfil)}+(q-2)\cdot h_{\yng(1)}\cdot z_{\young(\star,\hfil)}\cdot z_{\young(\hfil,\hfil)}\right)^b\\
\fpart_{\young(\hfil\hfil)}&=&\left( h_{\young(\star)}\cdot z_{\young(\star,\hfil)}^2+h_{\yng(1)}\cdot z_{\young(\hfil\hfil)}^2+(q-2)\cdot h_{\yng(1)}\cdot z_{\young(\hfil,\hfil)}^2\right)^b\\
\fpart_{\young(\hfil,\hfil)}&=&\left(h_{\young(\star)}\cdot z_{\young(\star,\hfil)}^2+2\cdot h_{\yng(1)}\cdot z_{\young(\hfil\hfil)}\cdot z_{\young(\hfil,\hfil)}+(q-3)\cdot h_{\yng(1)}\cdot z_{\young(\hfil,\hfil)}^2\right)^b
\label{dhlfield2}
\end{eqnarray}
\endnumparts
The renormalization map does not preserve the physical space, i.e. the image of \eref{initialDHL} (that is in general a submanifold of codimension 1) unless $q=2$. In fact, in such case the dynamical space is given by $[z_{\young(\star\star)}:z_{\young(\star,\hfil)}:z_{\young(\hfil\hfil)}]\in\proj^2$ and the map given by \eref{initialDHL} is surjective. One could in principle write the renormalization map in terms of Boltzmann weights associated to $\Jp,\Jnp,\Hy,\Hn$ (see e.g. \cite{zal,bly}); however, the map obtained in such variables is not rational (since it involves square roots) and its analysis is not as straightforward as it would be on the dynamical space. In any case it is convenient to perform the analysis in the dynamical space and then restrict to the physical space to obtain plots and thermodynamical quantities.\\
One can compute with good approximation the Green function of the renormalization map and look for phase transitions in the magnetic field part of the dynamical space. Looking just at the Ising case, with no surprise we find the full Lee-Yang circle for the ferromagnetic phase, and we obtain an anomalous plot for the supposedly paramagnetic phase (\fref{criticalphase} shows the $b=2$ case).
The anomalous plot illustrates two interesting facts. The first (proved in \cite{zal}) is that zeros of the partition function \emph{do} accumulate on the positive real axis even in the supposedly paramagnetic phase i.e. the system exhibits infinite susceptibility in the paramagnetic phase, which therefore is more appropriately called critical phase. The critical phase nevertheless exhibits paramagnetic behaviour (this is also proved in \cite{zal}); in fact, we report in figure \ref{magnetization} the numerical data for the spontaneous magnetization.
The second interesting fact to note is that points that are accumulating towards the positive real axis in the critical phase are not ordinary zeros of the partition function, but are preimages of the so-called indeterminacy set. In fact, the anomalous zeros in \fref{criticalphase} on the right, can be seen as $\times$s that decrease in size as they become dense, whereas the regular zeros in \fref{criticalphase} on the left form a solid line. The indeterminacy set is the set of points on which the renormalization map is not defined, i.e. the points that would map to all Boltzmann weights equal to $0$ under the renormalization map (see \cite{pap1}). Points accumulating on the positive real axis in \fref{criticalphase} correspond therefore to interactions that are only finitely renormalizable. Such points are in some sense anomalous from the points of view of both dynamics and physics, and it would be quite interesting to understand if this connection is more than just a mere coincidence. 
\section{Spider web}\label{spider}
\begin{figure}[!h]
\begin{center}
\begin{minipage}{3cm}
\includegraphics[height=2cm]{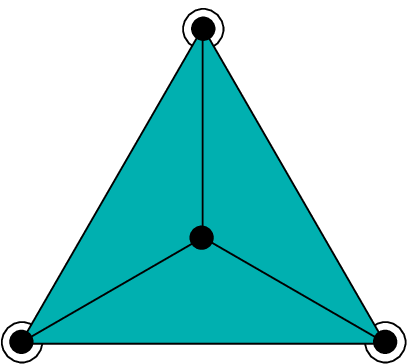}
\end{minipage}
\begin{minipage}{7.5cm}
\includegraphics[height=2.3cm]{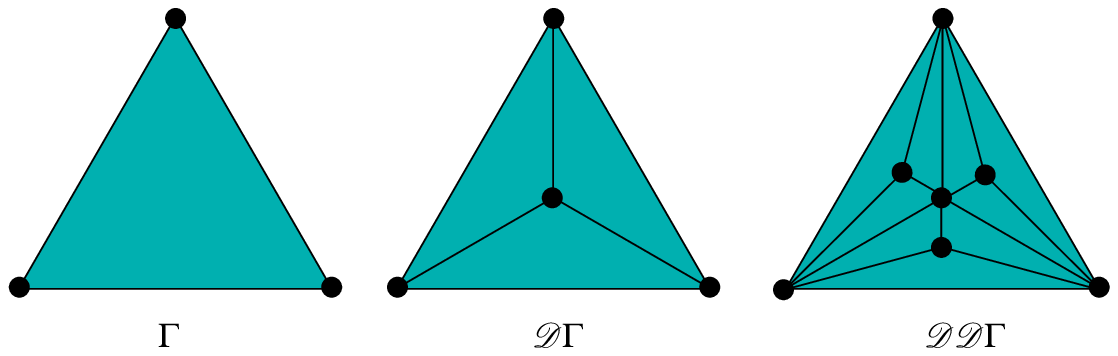}
\end{minipage}
\end{center}
\caption{Decoration for the spiderweb along with some iterations on a basic hypergraph}
\label{fig:spider}
\end{figure}

The spider web lattice is obtained by iterating the decoration $\deco$ shown in figure \fref{fig:spider} infinitely many times;
 as the picture illustrates this lattice is based on what is called a 3-uniform hypergraph. In this case the dynamical variables are the Boltzmann weights $[z_{\yng(3)}:z_{\yng(2,1)}:z_{\yng(1,1,1)}]\in\proj^2$ relative to the states of 3 neighbouring spins:
\begin{eqnarray*}
z_{\yng(3)}&\qquad{\rm same\ state }\\
z_{\yng(2,1)}&\qquad {\rm  two\ in\ the\ same\ state,\ third\ in\ different\ state}\\
z_{\yng(1,1,1)}&\qquad {\rm  three\ different\ states}
\end{eqnarray*}  
We can consider pair interactions given by $\Jp,\Jnp$ on each edge of each triangle in the following way:
\begin{equation}\fl
z_{\yng(3)}=\exp(-\beta\cdot3\Jp)\qquad z_{\yng(2,1)}=\exp(-\beta (\Jp+2\Jnp))\qquad z_{\yng(1,1,1)}=\exp(-\beta\cdot 3\Jnp)
\end{equation}
which follows by giving to each dynamical variable the Boltzmann weight associated to the energy of the pair interactions in the corresponding configuration. In this case, each side of each triangle (apart from the three sides of the initial hypergraph) is counted twice, as each side is shared by two 3-edges. Since this multiplicity is uniform for (almost) all sides, this is not an issue; the renormalization transformation is therefore easily written in the dynamical variables:
\numparts
\begin{eqnarray}
\fpart_{\yng(3)}&=& z_{\yng(3)}^3+(q-1)\cdot z_{\yng(2,1)}^3\\
\fpart_{\yng(2,1)}&=& z_{\yng(3)}\cdot z_{\yng(2,1)}^2+ z_{\yng(2,1)}^3+(q-2)\cdot z_{\yng(2,1)}\cdot z_{\yng(1,1,1)}^2\\
\fpart_{\yng(1,1,1)}&=&3\cdot z_{\yng(2,1)}^2\cdot z_{\yng(1,1,1)}+(q-3)\cdot z_{\yng(1,1,1)}^3
\end{eqnarray}
\label{spiderwebnof}
\endnumparts
Notice that, in general, the renormalization map does not preserve the physical space submanifold. Once more, this amounts to the well-known fact that in general renormalizing pair interactions gives rise to interactions that cannot be written as pair interactions. This did not happen in the previous case because the DHL is naturally defined using only 2-edges. Notice, moreover, that if $q=2$, the  equation for $z_{\yng(1,1,1)}$ uncouples from the first two and we have that the projective space generated by the first two variables is invariant under the renormalization map. This is not unexpected since, if $q=2$, there cannot be a configuration for which all three spins are in different states. As a matter of fact it is interesting to compute the restriction of the map in such a case, as we recover a map of the quadratic family best known as \emph{the cauliflower} (see for example \cite{dh}):
\[
\zeta^2=\eta\defeq\frac{z_{\yng(3)}}{z_{\yng(2,1)}}
\]\[
\eta\mapsto\frac{\fpart_{\yng(3)}}{\fpart_{\yng(2,1)}}=\frac{ z_{\yng(3)}^3+ z_{\yng(2,1)}^3}{ z_{\yng(2,1)}^2\left( z_{\yng(3)}+ z_{\yng(2,1)}\right)}=\frac{\eta^3+1}{\eta+1}=\eta^2-\eta+1.
\]
\begin{figure}[!htbp]
  \centering
  \includegraphics[width=3.5cm]{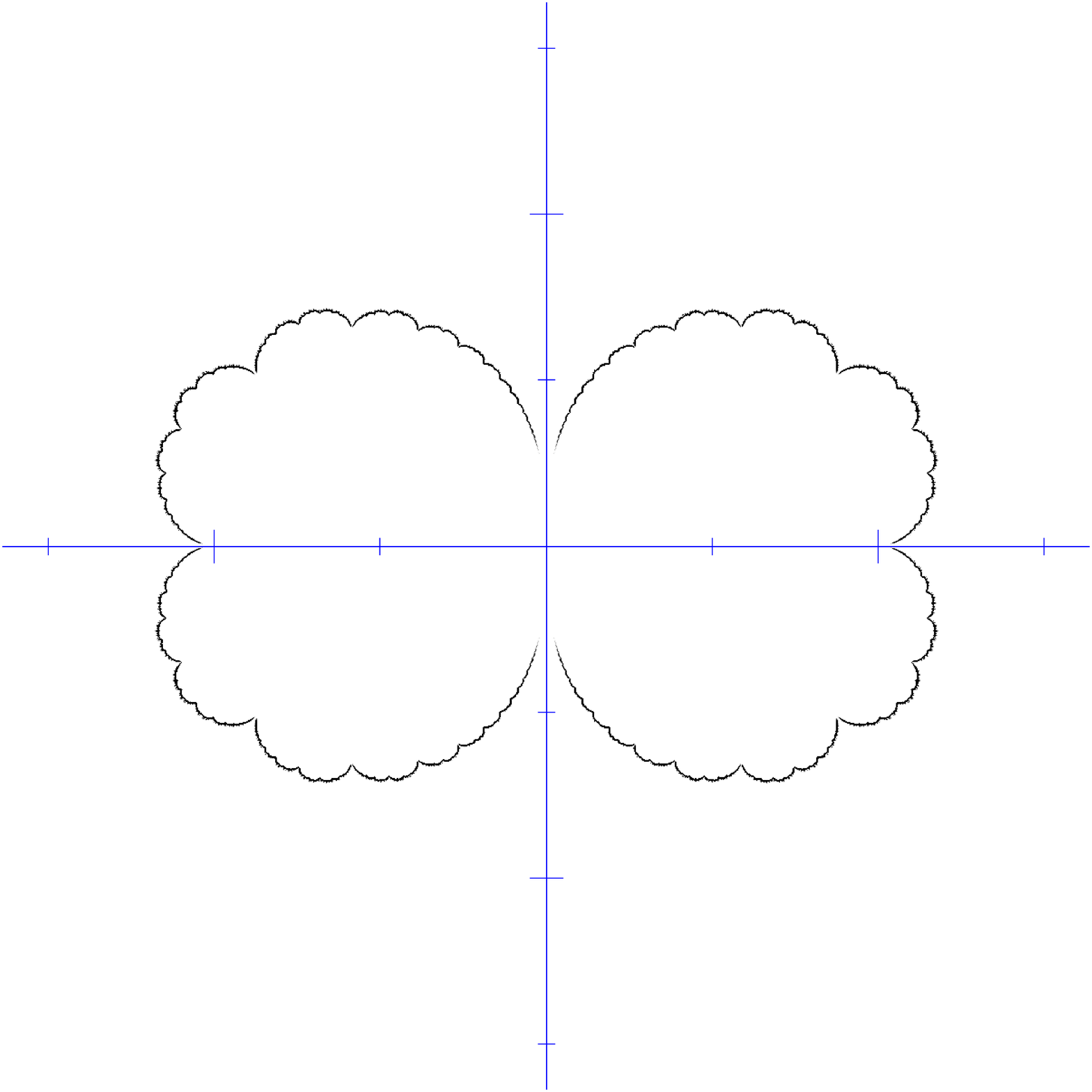}
  \includegraphics[width=3.5cm]{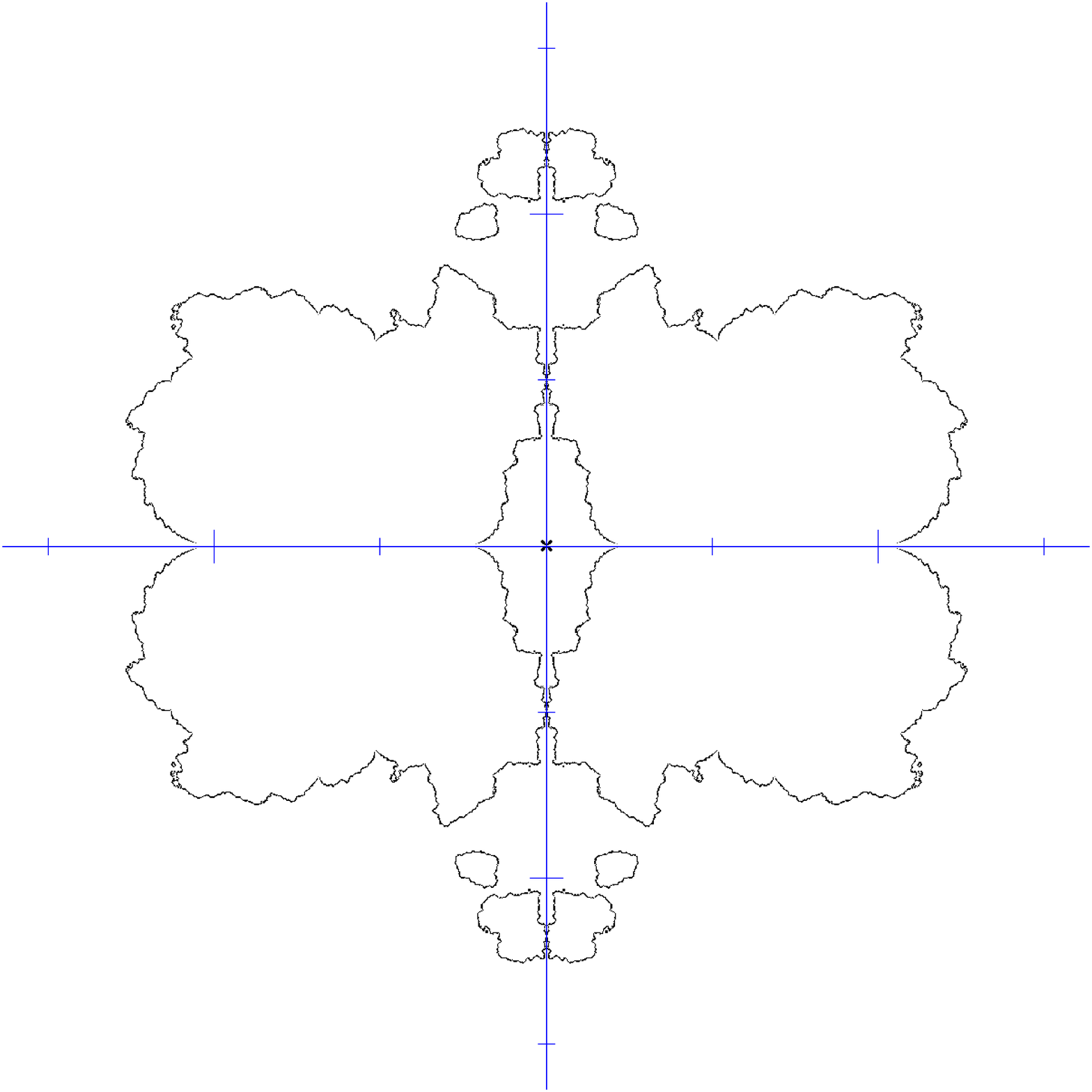}
  \includegraphics[width=3.5cm]{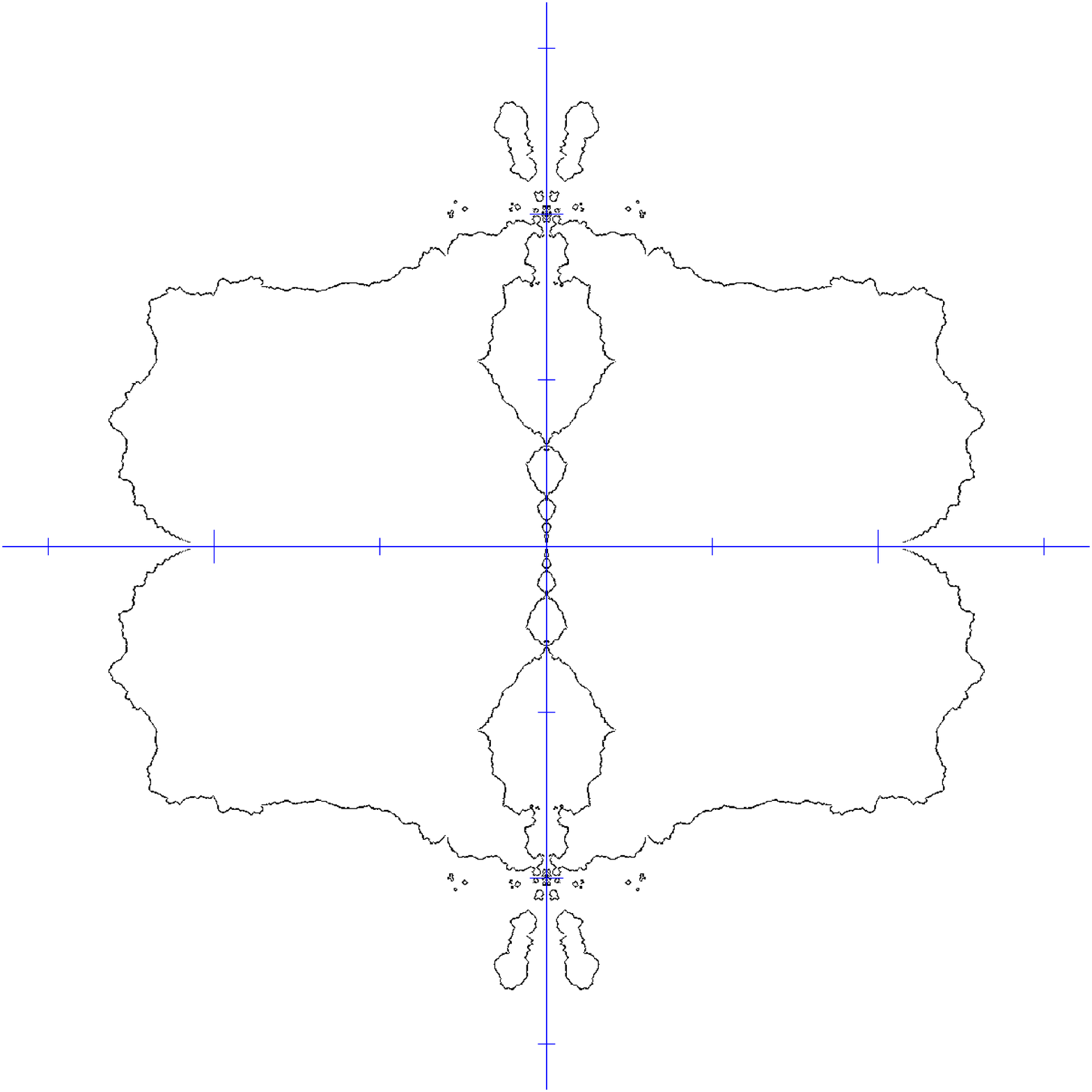}
\caption{\emph{Standard interaction chart}: Fisher zeros for $q=2,3,4$; as we change $q$ we can track the evolution of the antiferromagnetic phase.}
\label{ragnatela}
\end{figure}

We can recognize the cauliflower in the physical variables in the leftmost picture of figure \ref{ragnatela}. From the map we easily see that the paramagnetic point $[1:1]$ is a parabolic fixed point (i.e. its multiplier is a root of unity.) and the convergence of the Green function in its neighbourhood is rather slow. The same slow convergence rate can also be noticed for all preimages of this point . All points inside the cauliflower (therefore all antiferromagnetic interactions) will converge to the paramagnetic fixed point, while all points outside will converge to the ferromagnetic fixed point at infinity. This could be explained by the fact that frustration prevents the formation of an antiferromagnetic phase. 
For $q=3$ the map acts on the full $\proj^2$. This map has one indeterminacy point at $[0:0:1]$, represented by the cross at the center of the appropriate plot in figure \ref{ragnatela}. In fact, this point corresponds to an interaction that allows only for the configuration given by all three spins in different states. It is easy to check (see \fref{fig:spider}) that if $q=3$ it is not possible to have a configuration of spins on the spiderweb satisfying this requirement. To this extent, this interaction is not renormalizable. The newborn region that surrounds the indeterminacy point is mapped to the basin of attraction of the ferromagnetic point $\infty$, indicating that the behaviour of this phase could be antiferromagnetic. Considering $q=4$ or higher we observe that the antiferromagnetic phase disappears.\\
It is straightforward to write the renormalization group map in presence of an external magnetic field. However, for sake of clarity, we restrict ourselves to the case $q=2$; the dynamical variables are given by $[z_{\young(\star\star\star)}:z_{\young(\star\star,\hfil)}:z_{\young(\star,\hfil\hfil)}:z_{\yng(3)}]\in\proj^3$:
\numparts
\begin{eqnarray}
\fpart_{\young(\star\star\star)}&=& z_{\young(\star\star\star)}^3\cdot h_{\young(\star)}+ z_{\young(\star\star,\hfil)}^3\cdot h_{\young(\hfil)}\\
\fpart_{\young(\star\star,\hfil)}&=& z_{\young(\star\star\star)}\cdot z_{\young(\star\star,\hfil)}^2\cdot h_{\young(\star)}+ z_{\young(\star\star,\hfil)}\cdot z_{\young(\star,\hfil\hfil)}^2\cdot h_{\young(\hfil)}\\
\fpart_{\young(\star,\hfil\hfil)}&=& z_{\young(\star\star,\hfil)}^2 \cdot z_{\young(\star,\hfil\hfil)}\cdot h_{\young(\star)}+ z_{\young(\star,\hfil\hfil)}^2\cdot z_{\yng(3)}\cdot h_{\young(\hfil)}\\
\fpart_{\yng(3)}&=& z_{\young(\star,\hfil\hfil)}^3\cdot h_{\young(\star)}+ z_{\young(\hfil\hfil\hfil)}^3\cdot h_{\young(\hfil)}
\end{eqnarray}
\label{SpiderWebfield}
\endnumparts
\begin{figure}[!h]
\centering
  \includegraphics[height=5cm,angle=270]{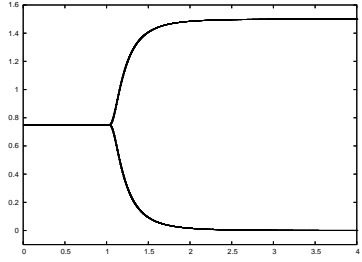}
\caption{Ising model on the spider web: spontaneous magnetization vs. $z$ variable for $z\in\reals$ for various values of interaction. Note that the transition is rather gentle; this is due to the fact that the density of zeros is low near the transition point}
\label{spidermagn} 
\end{figure}
Notice that the renormalization map is symmetric for the exchange of the special state with the other state. 
In figure \ref{spidermagn} we provide a plot of the spontaneous magnetization vs. interaction that confirms the presence of the paramagnetic phase for all antiferromagnetic interactions and of the ferromagnetic phase for all ferromagnetic interactions. 
\label{spiderweb}
\section{Sierpinski gasket}\label{sierpinski}
\begin{figure}[!h]    \begin{minipage}{3cm}
      \begin{center}
        \includegraphics[height=2.3cm]{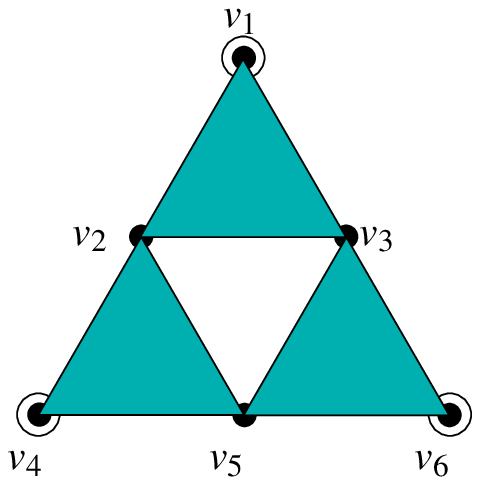}\\[0.8pt]
      \end{center}
    \end{minipage}
    \begin{minipage}{7.5cm}
      \begin{center}
        \includegraphics[height=2.3cm]{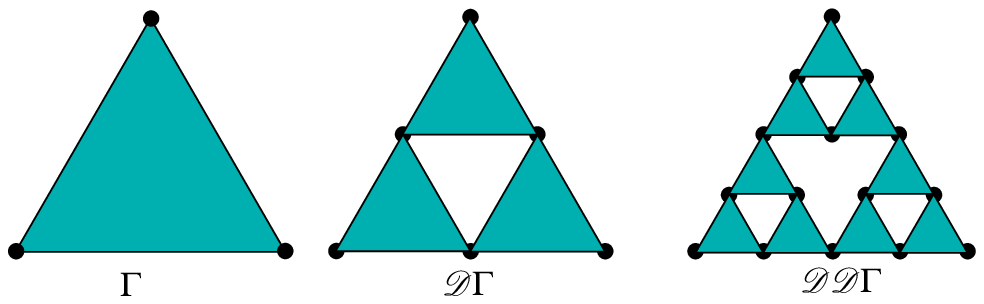}
      \end{center}
    \end{minipage}

\caption{Some iterations of the decoration which generates the Sierpinski gasket.}
\label{figure:sierpinski}
\end{figure}

We can generate the Sierpinski gasket by infinite iterations of the decoration shown in figure \ref{figure:sierpinski}. 

\begin{figure}[!htbp]
  \centering
  \includegraphics[width=5cm]{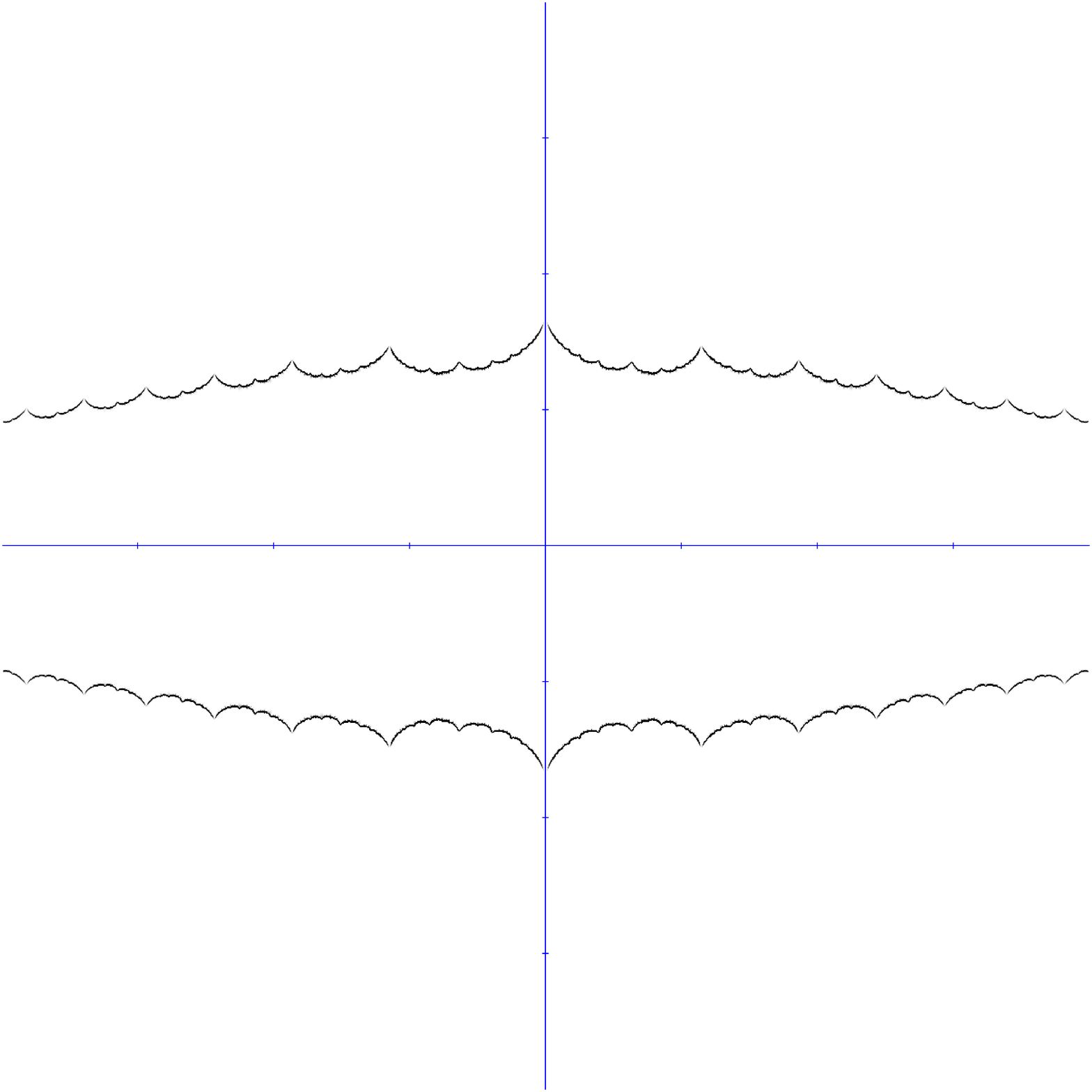}
  \includegraphics[width=5cm]{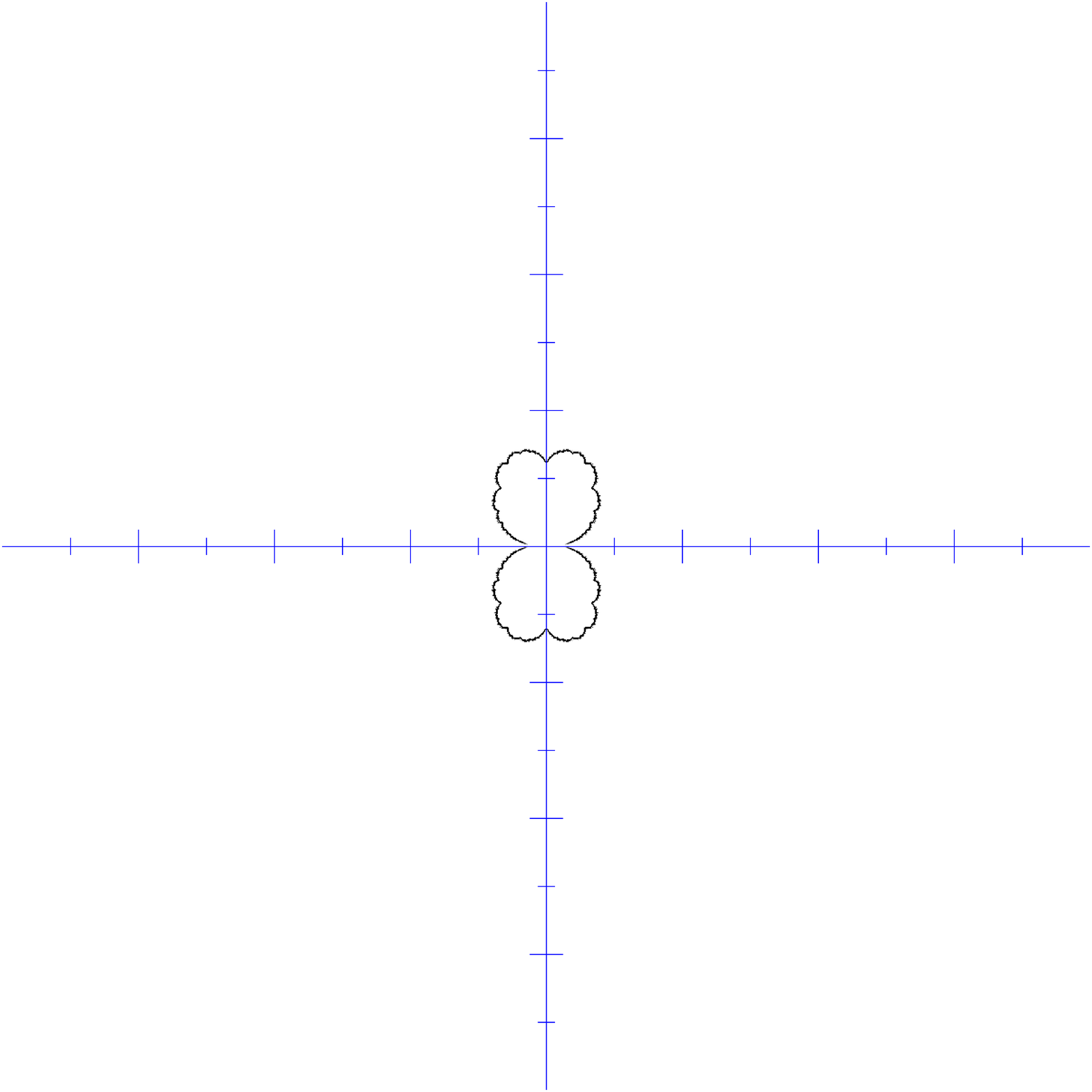} \\
  \includegraphics[width=5cm]{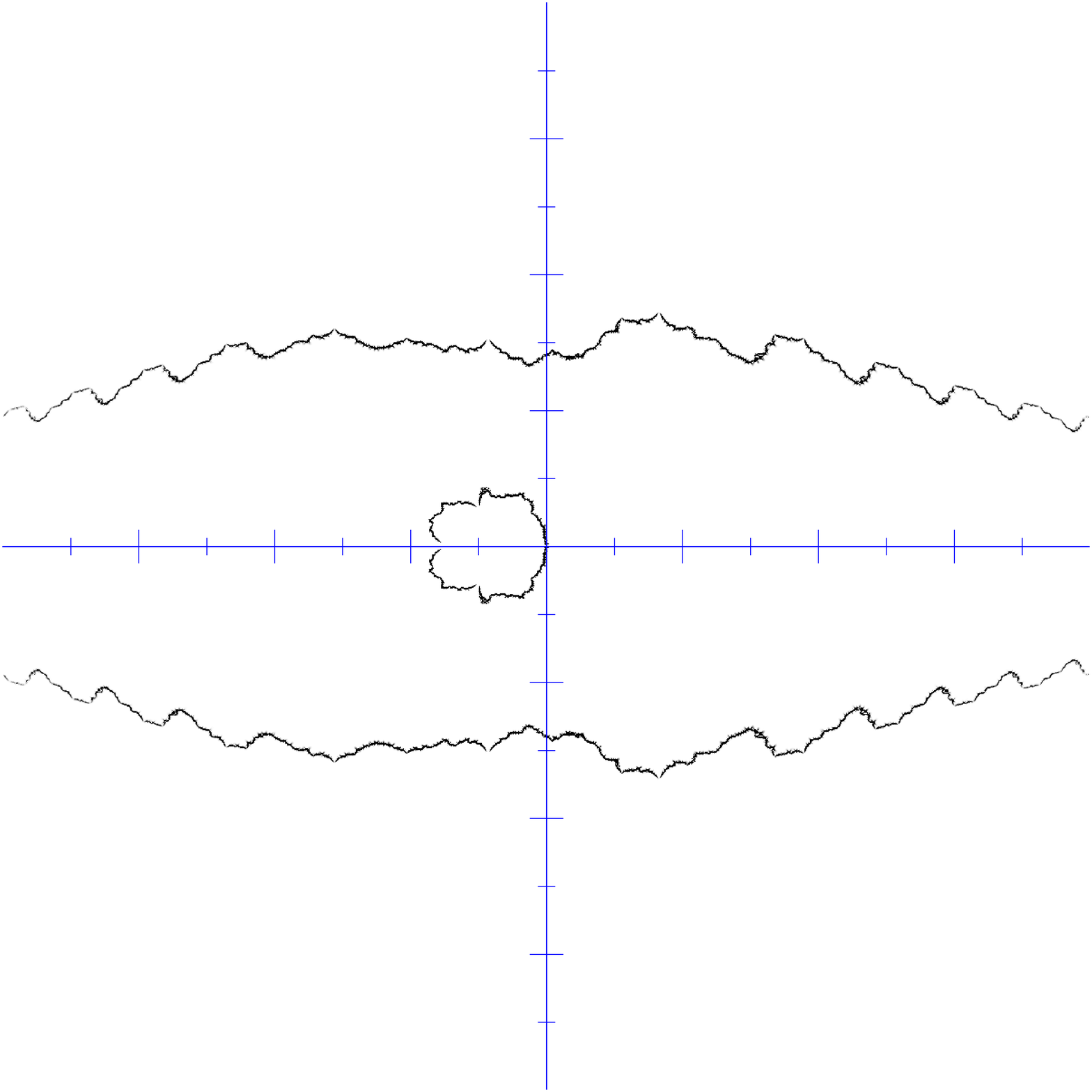}
  \includegraphics[width=5cm]{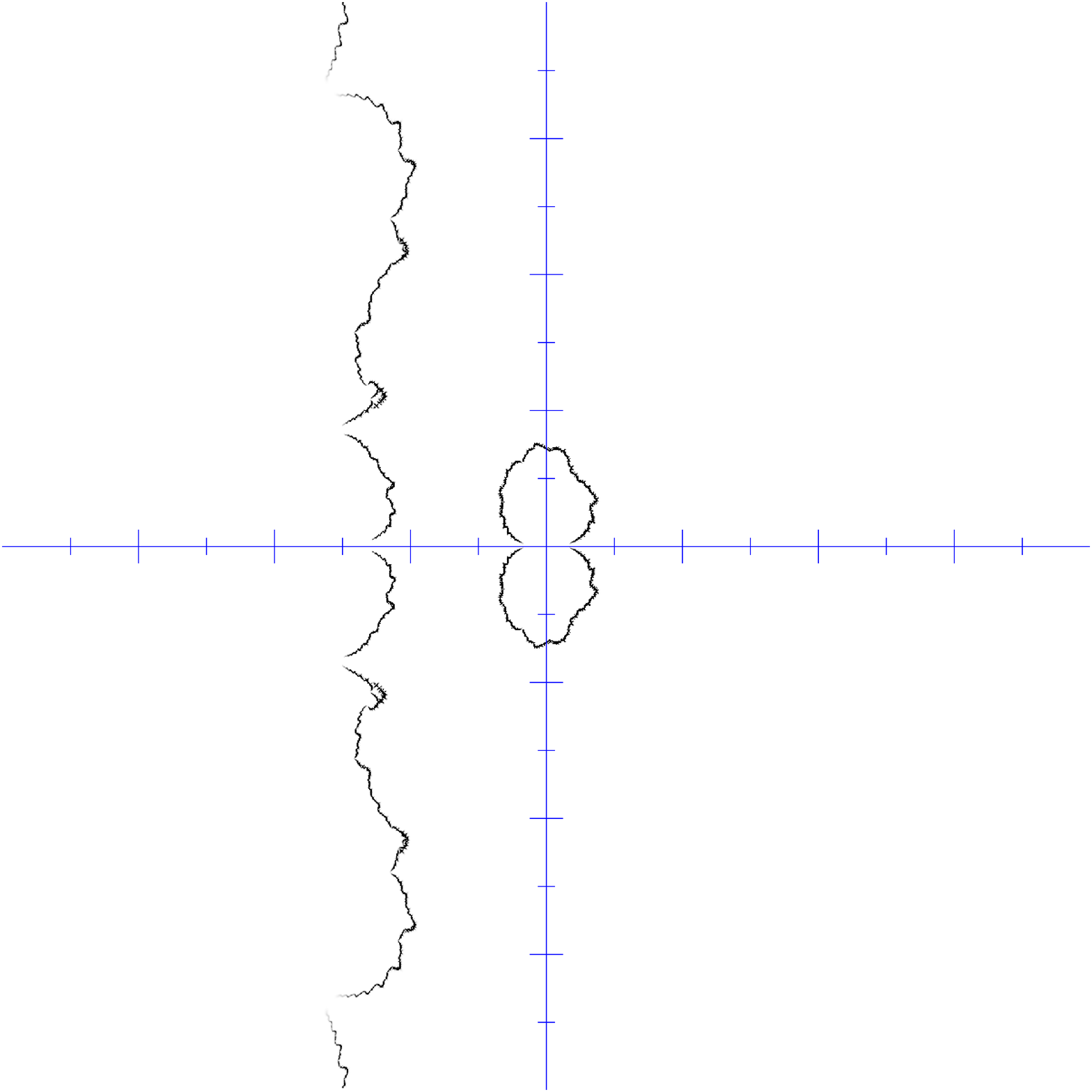}\\
  \includegraphics[width=5cm]{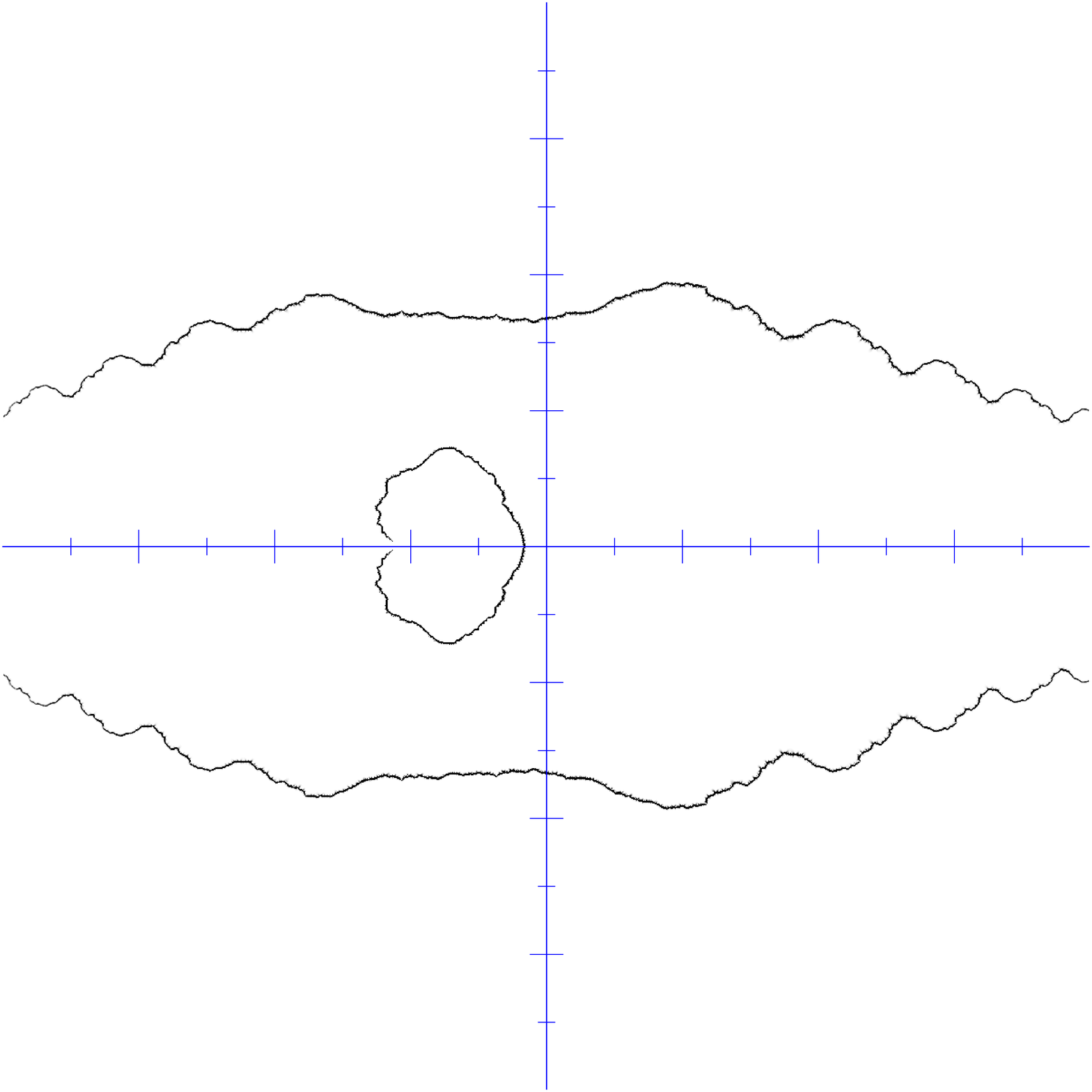}
  \includegraphics[width=5cm]{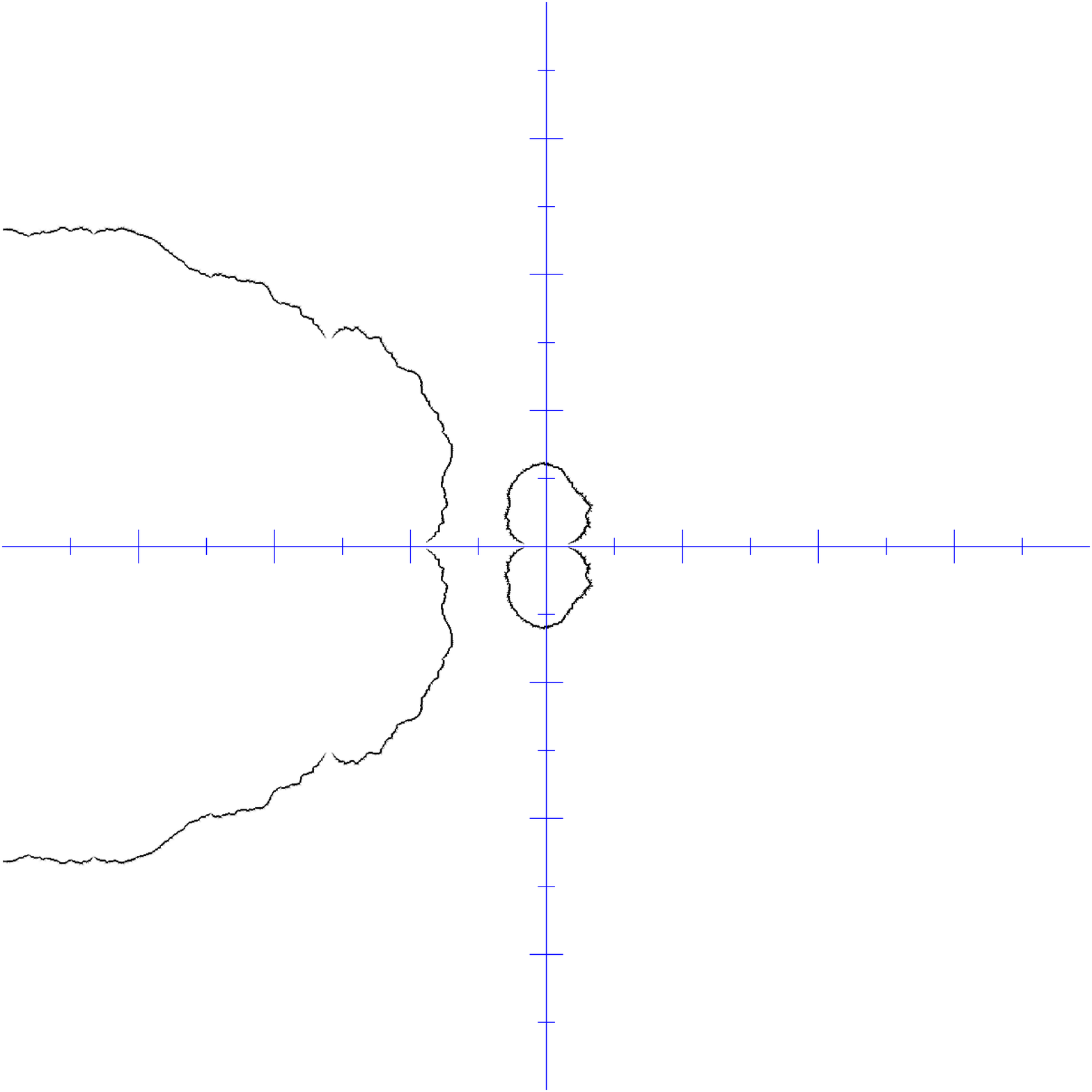}\\
\caption{Fisher zeros for the Sierpinski gasket. The left column shows zero
es in the interaction coordinates, while the right column show zeros for the inverse interaction coordinates; the
 three rows correspond to different $q=2,3,4$.}
\label{sierpiq}
\end{figure}

\begin{figure}[!h]
\centering
  \includegraphics[width=3.5cm]{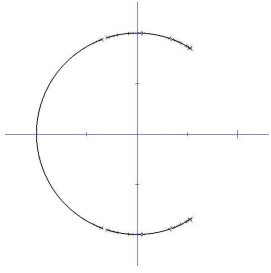}
\caption{Standard field coordinates: Lee-Yang zeros for the Sierpinski gasket. The picture is for $z=1.3$ but qualitatively depicts all ferromagnetic interactions. As we expected, zeros do not accumulate on the real positive axis, and their structure is quite complicated as it is made of pieces with genuine zeros and pieces with points of the indeterminacy set.}
\label{lysierp}
\end{figure}

The dynamical and physical spaces are the same as in \sref{spider}; in this case each side of each triangle is counted just once, so we have no multiplicity issues. We record for sake of completeness the renormalization map for any value of $q$:
\numparts
\label{sierpimap} 
\begin{equation}
\eqalign{
\fl
 \fpart_{\yng(3)}= z_{\yng(1,1,1)}^{3}\cdot \left(q-3\right)\*\left(q-2\right)\*\left(q-1\right)+3\cdot z_{\yng(2,1)}\cdot  z_{\yng(1,1,1)}^{2}\cdot \left(q-2\right)\*\left(q-1\right)+\cr z_{\yng(2,1)}^{3}\cdot \left(q-1\right)+3\cdot z_{\yng(2,1)}^{2}\cdot  z_{\yng(1,1,1)}\cdot \left(q-2\right)\*\left(q-1\right)+\cr+ 3\cdot z_{\yng(2,1)}^{3}\cdot \left(q-1\right)+3\cdot z_{\yng(3)}\cdot  z_{\yng(2,1)}^{2}\cdot \left(q-1\right)+ z_{\yng(3)}^{3}}
\end{equation}
\begin{equation}
\eqalign{
\fl
\fpart_{\yng(2,1)}= z_{\yng(1,1,1)}^{3}\cdot \left(q-4\right)\*\left(q-3\right)\*\left(q-2\right)+\left( z_{\yng(1,1,1)}^{3}+2\cdot  z_{\yng(2,1)}\cdot  z_{\yng(1,1,1)}^{2}\right)\*\left(q-3\right)\*\left(q-2\right)+\cr+\left(2\cdot  z_{\yng(2,1)}\cdot  z_{\yng(1,1,1)}^{2}+ z_{\yng(2,1)}^{2}\cdot  z_{\yng(1,1,1)}\right)\*\left(q-3\right)\*\left(q-2\right)+\cr+3\cdot  z_{\yng(2,1)}\cdot  z_{\yng(1,1,1)}^{2}\cdot \left(q-3\right)\*\left(q-2\right)+\left( z_{\yng(2,1)}\cdot  z_{\yng(1,1,1)}^{2}+2\cdot  z_{\yng(2,1)}^{2}\cdot  z_{\yng(1,1,1)}\right)\*\left(q-2\right)+\cr+\left(2\cdot  z_{\yng(2,1)}\cdot  z_{\yng(1,1,1)}^{2}+2\cdot  z_{\yng(2,1)}^{2}\cdot  z_{\yng(1,1,1)}+2\cdot  z_{\yng(2,1)}^{3}\right)\*\left(q-2\right)+\cr+\left( z_{\yng(3)}\cdot  z_{\yng(1,1,1)}^{2}+2\cdot  z_{\yng(2,1)}^{2}\cdot  z_{\yng(1,1,1)}\right)\*\left(q-2\right)+\cr+\left(2\cdot  z_{\yng(2,1)}^{2}\cdot  z_{\yng(1,1,1)}+ z_{\yng(2,1)}^{3}\right)\*\left(q-2\right)+\cr+\left(2\cdot  z_{\yng(3)}\cdot  z_{\yng(2,1)}\cdot  z_{\yng(1,1,1)}+ z_{\yng(2,1)}^{3}\right)\*\left(q-2\right)+ z_{\yng(2,1)}^{3}\cdot \left(q-2\right)+\cr+\left(2\cdot  z_{\yng(2,1)}^{3}+ z_{\yng(3)}\cdot  z_{\yng(2,1)}^{2}\right)+\left( z_{\yng(2,1)}^{3}+2\cdot  z_{\yng(3)}\cdot  z_{\yng(2,1)}^{2}\right)+\cr+ z_{\yng(3)}\cdot  z_{\yng(2,1)}^{2}+ z_{\yng(3)}^{2}\cdot  z_{\yng(2,1)}}
\end{equation}
\begin{equation}
\eqalign{
\fl
\fpart_{\yng(1,1,1)}= z_{\yng(1,1,1)}^{3}\cdot \left(q-5\right)\*\left(q-4\right)\*\left(q-3\right)+3\*\left( z_{\yng(1,1,1)}^{3}+2\cdot  z_{\yng(2,1)}\cdot  z_{\yng(1,1,1)}^{2}\right)\*\left(q-4\right)\*\left(q-3\right)+\cr+3\cdot  z_{\yng(2,1)}\cdot  z_{\yng(1,1,1)}^{2}\cdot \left(q-4\right)\*\left(q-3\right)+\cr+3\*\left( z_{\yng(1,1,1)}^{3}+2\cdot  z_{\yng(2,1)}\cdot  z_{\yng(1,1,1)}^{2}+3\cdot  z_{\yng(2,1)}^{2}\cdot  z_{\yng(1,1,1)}\right)\*\left(q-3\right)+\cr+3\*\left( z_{\yng(2,1)}\cdot  z_{\yng(1,1,1)}^{2}+2\cdot  z_{\yng(2,1)}^{2}\cdot  z_{\yng(1,1,1)}\right)\*\left(q-3\right)+\cr+3\*\left( z_{\yng(3)}\cdot  z_{\yng(1,1,1)}^{2}+2\cdot  z_{\yng(2,1)}^{2}\cdot  z_{\yng(1,1,1)}\right)\*\left(q-3\right)+ z_{\yng(2,1)}^{3}\cdot \left(q-3\right)+\cr+ z_{\yng(1,1,1)}^{3}+9\cdot z_{\yng(2,1)}^{2}\cdot  z_{\yng(1,1,1)}+6\cdot  z_{\yng(3)}\cdot  z_{\yng(2,1)}\cdot  z_{\yng(1,1,1)}+\cr+8\cdot  z_{\yng(2,1)}^{3}+3\cdot  z_{\yng(3)}\cdot  z_{\yng(2,1)}^{2}}
\end{equation}
\endnumparts
Once more, if $q=2$, the third equation decouples and again we obtain a map defined on a $\proj^1$.
\numparts
\begin{eqnarray}
 \fpart_{\ytiny\yng(3)}&=&4\cdot  z_{\yng(2,1)}^{2}- z_{\yng(3)}\cdot  z_{\yng(2,1)}+ z_{\yng(3)}^{2}\\
\fpart_{\ytiny\yng(2,1)}&=& z_{\yng(2,1)}\cdot\left(3\cdot  z_{\yng(2,1)}+ z_{\yng(3)}\right).
\end{eqnarray}
\endnumparts
This is equation 3.2 in \cite{gasm} or equation 14 in \cite{bcd}. The exact and numerical results (\fref{sierpiq}) tell us that we have no phase transitions at finite temperature; we have zeros in the thermodynamical limit only for $T=0$ in the physical domain. The paramagnetic fixed point is attracting for all points in the positive real axis, so that we cannot have a ferromagnetic phase. This behaviour is similar to that of the linear chain.
Again it is easy but tedious to compute the map for the general case in presence of an external magnetic field; we will give here the exact expression for the Ising case:
\numparts
\begin{eqnarray}
\fpart_{\young(\star\star\star)}&=& h_{\young(\star)}^3\cdot z_{\young(\star\star\star)}^3+3\cdot h_{\young(\star)}^2h_{\yng(1)}\cdot \left( z_{\young(\star\star\star)}\cdot z_{\young(\star\star,\hfil)}^2\right)+\nonumber\\&&+3\cdot h_{\young(\star)}h_{\yng(1)}^2\cdot\left( z_{\young(\star,\hfil\hfil)}\cdot z_{\young(\star\star,\hfil)}^2\right)+h_{\yng(1)}^3\cdot z_{\young(\star,\hfil\hfil)}^3\\
\fpart_{\young(\star\star,\hfil)}&=&  h_{\young(\star)}^3\cdot z_{\young(\star\star\star)}^2\cdot z_{\young(\star\star,\hfil)}+ h_{\young(\star)}^2h_{\yng(1)}\cdot\left( z_{\young(\star\star,\hfil)}^3+2\cdot z_{\young(\star,\hfil\hfil)}\cdot z_{\young(\star\star\star)}\cdot z_{\young(\star\star,\hfil)}\right)+\nonumber\\&&+h_{\young(\star)}h_{\yng(1)}^2\cdot\left( z_{\young(\star\star,\hfil)}^2\cdot z_{\young(\hfil\hfil\hfil)}+2\cdot z_{\young(\star,\hfil\hfil)}^2\cdot z_{\young(\star\star,\hfil)}\right)+ h_{\yng(1)}^3\cdot z_{\young(\hfil\hfil\hfil)}\cdot z_{\young(\star,\hfil\hfil)}^2\\
\fpart_{\young(\star,\hfil\hfil)}&=&  h_{\yng(1)}^3\cdot z_{\young(\hfil\hfil\hfil)}^2\cdot z_{\young(\star,\hfil\hfil)}+h_{\young(\star)}h_{\yng(1)}^2\cdot\left( z_{\young(\star,\hfil\hfil)}^3+2\cdot z_{\young(\star\star,\hfil)}\cdot z_{\young(\hfil\hfil\hfil)}\cdot z_{\young(\star,\hfil\hfil)}\right)+\nonumber\\&&+ h_{\young(\star)}^2h_{\yng(1)}\cdot\left( z_{\young(\star,\hfil\hfil)}^2\cdot z_{\young(\star\star\star)}+2\cdot z_{\young(\star\star,\hfil)}^2\cdot z_{\young(\star,\hfil\hfil)}\right)+ h_{\young(\star)}^3\cdot z_{\young(\star\star\star)}\cdot z_{\young(\star\star,\hfil)}^2\\
\fpart_{\young(\hfil\hfil\hfil)}&=& h_{\yng(1)}^3\cdot z_{\young(\hfil\hfil\hfil)}^3+3\cdot h_{\young(\star)}h_{\yng(1)}^2\cdot\left( z_{\young(\hfil\hfil\hfil)}\cdot z_{\young(\star,\hfil\hfil)}^2\right)+\nonumber\\&&+3\cdot h_{\young(\star)}^2h_{\yng(1)} \cdot\left( z_{\young(\star\star,\hfil)}\cdot z_{\young(\star,\hfil\hfil)}^2\right)+ h_{\young(\star)}^3\cdot z_{\young(\star\star,\hfil)}^3
\end{eqnarray}
\endnumparts
Notice that, once again, we have complete symmetry for exchange of the special state with the other. Since we have no ferromagnetic phase, the Lee-Yang zeros do not accumulate to the positive real axis, as shown in figure \ref{lysierp}.
\section{Cylinders}\label{pipes} 
\begin{figure}[!htbp]
  \centering
  \includegraphics[height=2cm]{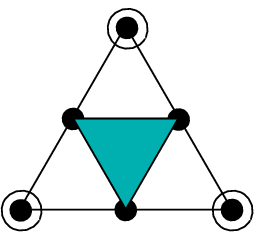}
  \includegraphics[height=2cm]{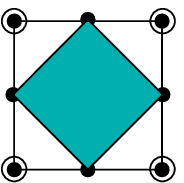}
  \includegraphics[height=2cm]{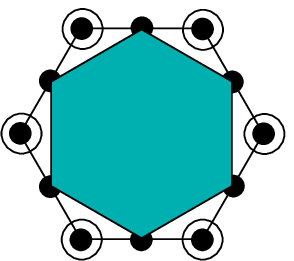}
  \caption{Decorations that generate the skewed cylinders $\mathbb{Z}^2/r\cdot\mathbb{Z}(1,1)$ for $r=3,4,6$, respectively. The cylinders are obtained by substituting infinitely many times the triangle (square, hexagon) with the corresponding decoration}
  \label{fig:decotubo}
\end{figure}

\begin{figure}[!htbp]
  \centering
  \includegraphics[height=2cm]{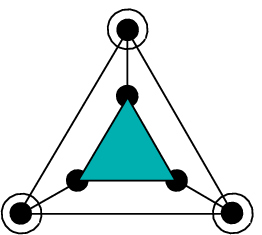}
  \includegraphics[height=2cm]{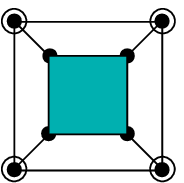}
  \includegraphics[height=2cm]{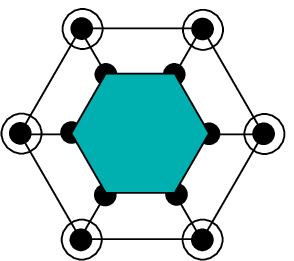}
\caption{Decorations that generate the cylinders $\mathbb{Z}^2/r\cdot\mathbb{Z}(1,0)$ for $r=3,4,6$, respectively. The cylinders are obtained by substituting infinitely many times the triangle (square, hexagon) with the corresponding decoration}
  \label{fig:decotubo1}
\end{figure}

In this final section we provide an example of non-uniform lattices, i.e. lattices in which several types of edges are used. We present a lattice obtained as the quotient of the square lattice $\mathbb{Z}^2$ with a translation. Such lattices can be regarded as being generated by decorations in \fref{fig:decotubo} and \fref{fig:decotubo1}. For these lattices, a very special case of non-uniform lattices, we recover results that can be found in a completely equivalent way using the transfer matrix method; in this framework the transfer matrix is indeed the renormalization map.\\
In each one of these lattices we have two type of edges. One is a regular 2-edge and the other is respectively a 3,4 or 6-edge. Consider, for instance, the simplest lattice in \fref{fig:decotubo}, i.e. the 3-skewed cylinder and let $q=2$. The dynamical variables are the Boltzmann weights $[z_{\ytiny\yng(3)}:z_{\ytiny\yng(2,1)}],[z_{\ytiny\yng(2)}:z_{\ytiny\yng(1,1)}]\in\proj^1\times\proj^1$. The renormalization map will leave pair interactions (i.e. the second factor) invariant and will induce a 3-spin interaction on the first factor according to the following formula:
\numparts 
\begin{eqnarray}
\fpart_{\ytiny\yng(3)}&=& z_{\yng(3)}\cdot z_{\yng(2)}^6+3\cdot z_{\yng(2,1)}\left( z_{\yng(2)}^4\cdot z_{\yng(1,1)}^2+ z_{\yng(2)}^2\cdot z_{\yng(1,1)}^4\right)+\nonumber\\&&+ z_{\yng(3)}\cdot z_{\yng(1,1)}^6\\
\fpart_{\ytiny\yng(2,1)}&=& z_{\yng(3)}\cdot z_{\yng(2)}^4\cdot z_{\yng(1,1)}^2+3\cdot z_{\yng(2,1)}\left( z_{\yng(2)}^4\cdot z_{\yng(1,1)}^2+ z_{\yng(2)}^2\cdot z_{\yng(1,1)}^4\right)+\nonumber\\&&+ z_{\yng(3)}\cdot z_{\yng(2)}^2\cdot z_{\yng(1,1)}^4\\[4pt]
\fpart_{\ytiny\yng(2)}&=& z_{\yng(2)}\\
\fpart_{\ytiny\yng(1,1)}&=& z_{\yng(1,1)}
\end{eqnarray}
\label{tubo3}
\endnumparts
Notice that we can arrange the map as a linear map in the order 3 variables, parametric in the order 2 variables:
\[
\left(
\begin{array}{c}
\fpart_{\ytiny\yng(3)}\\\fpart_{\ytiny\yng(2,1)}
\end{array}
\right)=
\left(\begin{array}{cc}
 z_{\yng(2)}^6+ z_{\yng(1,1)}^6&3\left( z_{\yng(2)}^4\cdot z_{\yng(1,1)}^2+ z_{\yng(2)}^2\cdot z_{\yng(1,1)}^4\right)\\
\left( z_{\yng(2)}^4\cdot z_{\yng(1,1)}^2+ z_{\yng(2)}^2\cdot z_{\yng(1,1)}^4\right)&3\left( z_{\yng(2)}^4\cdot z_{\yng(1,1)}^2+ z_{\yng(2)}^2\cdot z_{\yng(1,1)}^4\right)
\end{array}\right)
\left(
\begin{array}{c}
z_{\ytiny\yng(3)}\\z_{\ytiny\yng(2,1)}
\end{array}
\right),
\]
whose corresponding matrix is the transfer matrix of the system.
Since we are dealing with a projective space we can factor out the polynomial $z_{\yng(2)}^4\cdot z_{\yng(1,1)}^2+ z_{\yng(2)}^2\cdot z_{\yng(1,1)}^4$ (if  $z^2_{\yng(2)}+z^2_{\yng(1,1)}\not=0$), 
and defining:
\[
a\left(\left[ z_{\yng(2)}: z_{\yng(1,1)}\right]\right)\defeq\frac{ z_{\yng(2)}^6+ z_{\yng(1,1)}^6}{\left( z_{\yng(2)}^4\cdot z_{\yng(1,1)}^2+ z_{\yng(2)}^2\cdot z_{\yng(1,1)}^4\right)},
\]
we can rewrite the matrix in the much simpler form:
\[
\left(
\begin{array}{cc}
a&3\\1&3
\end{array}
\right).
\]
Computing $a$ in the standard interaction coordinates $\zeta$ (for $\zeta\not =\pm{\rm i})$, we obtain:
\[
\zeta \defeq\left(\frac{z_{\yng(2)}}{z_{\yng(1,1)}}\right)\quad
a=\frac{\zeta^6+1}{\zeta^2(\zeta^2+1)}=\frac{\zeta^4-\zeta^2+1}{\zeta^2}.
\]
 \begin{figure}[!!h]
  \centering
\includegraphics[width=5cm]{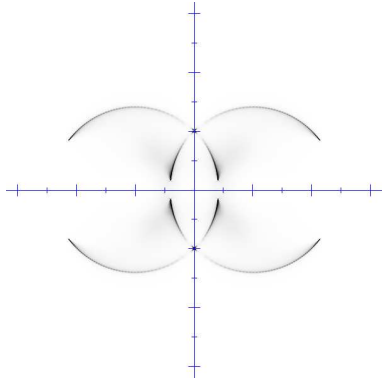}
  \caption{\emph{Standard interaction coordinates}: The set of non-analyticity points of the Green function; it looks like a subset of the Fisher zero set of the $\mathbb{Z}^2$ lattice.}
  \label{fig:euro}
\end{figure}

We can compute the Green function in the variable $\zeta$ and obtain the set depicted in figure \ref{fig:euro} for the non-analyticity locus.\\
Notice that in this case the matrix of degrees does not satisfy the Perron-Frobenius hypothesis of \ref{green}; in fact, the matrix is parabolic, i.e. it is not diagonalizable, with generalized eigenvalue 1. Therefore, we have to use a variation of the argument that we provided; the convergence of the Green function in this case is much slower (logarithmic) and the plot looks less definite.  
Indeed, one can obtain the set in an analytical way; in fact, one can easily check that in this case the appropriate version of the Green function is proportional to the logarithm of the norm of the maximal eigenvalue of the matrix. The non-analyticity locus is therefore contained in the set of points such that we have at least two eigenvalues with maximum norm. Such a condition is easily expressed in an analytic form and the resulting set agrees with the numerical picture.\\
\ack
It is my special pleasure to thank Prof. Stefano Marmi for his constant support and always welcome suggestions. Also I am indebted to anonymous referees for their most useful comments.   
\appendix
\section{Numerical study of rational maps}
\subsection{Fixed points and basins of attraction}\label{basins} 
In cases where it is easy to locate all stable fixed points of the map (e.g. for maps on the Riemann sphere $\smash{\hat\cplx}$), it is possible to obtain their basins of attraction in the following way. First of all we find a stable neighbourhood of each fixed point i.e. a ball of small radius such that its image is contained in itself. Having fixed a maximum number of iterations, we discretize a bounded region of the \emph{physical space} in a finite number of \emph{pixels}, and for each pixel, we apply iteratively the map starting from the center of the pixel until we end up in one of the aforementioned stable neighbourhoods. If this happens in the given maximum number of iterations, we declare the initial pixel to belong to the \emph{attracting basin} of the corresponding fixed point and we color it according to the speed of convergence (the faster the lighter) and to which fixed point it is attracted to. If the point does not fall in any neighbourhood in the given number of iterations, it is coloured black.\\
Pictures obtained in such a way show the unstable set of the map as the boundary of the basins of attraction; moreover, showing which points are attracted to which fixed points, the pictures provide basic information on the asymptotic dynamics of the RG map.  
\subsection{Green Function}\label{green}
We recall that a rational map on a multiprojective space $\dynspace$ lifts to a polynomial map that is separately homogeneous in each factor, i.e.:
\[
\fl
\begin{array}{rl}
f:&\proj^{r_1}\times\cdots\times\proj^{r_p}\to\proj^{r_1}\times\cdots\times\proj^{r_p}\\
  &\left([z^{(1)}],\cdots,[z^{(p)}]\right)\mapsto\left(\left[F^{(1)}\left(z^{(1)},\cdots,z^{(p)}\right)\right],\cdots,[F^{(p)}\left(z^{(1)},\cdots,z^{(p)}\right)]\right)
\end{array}
\]
where each $F^{(i)}$ is such that:
\[
F^{(i)}\left(\nu_1 z^{(1)},\cdots,\nu_p z^{(p)}\right)=\prod_j \nu_j^{d_j^i}F^{(i)}\left(z^{(1)},\cdots,z^{(p)}\right)
\]
and $d_j^i$ is the degree of $F^{(i)}$ with respect to $z^{(j)}$.
Considering $d_j^i$ as an integer-valued matrix $D$, we can find its eigenvalues; in good cases we expect (via the Perron-Frobenius theorem) a simple real maximal eigenvalue $\rho_+>1$ such that its associated (normalized) eigenvector $w_+$ has all non-negative coordinates. In such cases we can define the Green function:
\[
\mathscr{G}=\lim_{n\to\infty}\frac{1}{\rho_+^n}\left\langle w_+,\log\left\|F^n\left(z^{(1)},\cdots,z^{(p)}\right)\right\|\right\rangle
\]
Notice the similarity of this function with the free energy of the system. In fact, for hierarchical lattices we have the following expression for the free energy:
\[
\enel=\lim_{n\to\infty}\frac{1}{\# {\rm{edges\ }} \Gamma_n}\log\left\|\fpart_0\left(F^n\left(z^{(1)},\cdots,z^{(p)}\right)\right)\right\|,
\]
where $\fpart_0$ is the partition function of the starting hypergraph. Moreover, if we call $\delta_i$ the number of edges of type $i$ that belong to the starting hypergraph, we can express the total number of edges of the $n$-th approximation to the hierarchical lattice as $\sum_{i,j}\delta_i \left(D^n\right)^i_j$. For generic $\delta$, this expression is obviously asymptotic to $\rho_+^n$. 
As explained in \cite{pap1} there are results that state that the two functions $\mathscr{G}$ and $\enel$ are equal in the uniform case with mild assumptions on $\fpart_0$, but there is no general result for the non-uniform case. Moreover notice that in the uniform case the matrix $D$ is just a number, therefore most of the computations are made easier.\\
We remark that we can exploit the homogeneous nature of the map to obtain a clever (and geometrically converging) way of numerically computing the Green function. In fact, let us define the sequence $z_n$ of normalized iterates and the sequence $\lambda_n$ of the corresponding norms as follows: 
\begin{eqnarray*}
\fl
\lambda_0^i\defeq\|z^{(i)}\|,&\qquad z_0^{(i)}\defeq z^{(i)}/\lambda_0^i\\
\fl
\lambda_{n+1}^i\defeq\left\|F^{(i)}\left(z_{n}^{(1)},z_{n}^{(2)},\cdots,z_{n}^{(p)}\right)\right\|,&\qquad z_{n+1}^{(i)}\defeq F^{(i)}\left(z_{n}^{(1)},z_{n}^{(2)},\cdots,z_{n}^{(p)}\right)/\lambda_{n+1}^i
\end{eqnarray*}
we can write:
\begin{eqnarray*}
\left\|F^{(i)}\left(z^{(1)},\cdots,z^{(p)}\right)\right\|&=&\prod_j \left(\lambda_0^j\right)^{d_j^i}\cdot \left\|F^{(i)}\left(z_0^{(1)},\cdots,z_0^{(p)}\right)\right\|\\
&=& \lambda_1^i\cdot\prod_j \left(\lambda_0^j\right)^{d_j^i}.
\end{eqnarray*}
Therefore, iterating the previous expression we get:
\begin{equation*}
\left\|\left(F^n\right)^{(i_n)}\left(\left\{z^{(k)}\right\}\right)\right\|=\lambda_n^{i_n}\cdot\prod_{k=0}^{n-1}\prod_{i_k\cdots i_{n-1}}\left(\lambda^{i_k}_k\right)^{d^{i_n}_{i_{n-1}}d^{i_{n-1}}_{i_{n-2}} \cdots d^{i_{k+1}}_{i_k}}.
\end{equation*}
Taking the logarithm and considering $\log\lambda^i_n$ as components of a vector $\log\blambda_n$ in a $p$-dimensional space and again $d_j^i$ as elements of the $p\times p$-matrix $D$, we obtain the following expression:
\[ \fl
\log\left\|\left(F^n\right)\left(\left\{z^{(k)}\right\}\right)\right\|=\log\blambda_n+D\log\blambda_{n-1}+D^2\log\blambda_{n-2}+\cdots+ D^n\log\blambda_0.\]
When we compute the scalar product with the maximal eigenvector of $D$ we are projecting on the corresponding eigenspace, therefore the expression can be rewritten as:
\begin{eqnarray*}
\left\langle w_+,\log\blambda_n\right\rangle+\left\langle w_+,D\log\blambda_{n-1}\right\rangle+\cdots+ \left\langle w_+,D^n\log\blambda_0\right\rangle=\\
=\left\langle w_+,\log\blambda_n\right\rangle+\rho_+\left\langle w_+,\log\blambda_{n-1}\right\rangle+\cdots+\rho_+^n\left\langle w_+,\log\blambda_{0}\right\rangle.
\end{eqnarray*}
Dividing by the normalization term we get the following expression for the Green function:
\[
\mathscr{G}=\lim_{N\to\infty}\sum_{n=0}^N\frac{\left\langle w_+,\log\blambda_n\right\rangle}{\rho_+^n}, 
\]
that is geometrically convergent (if $\rho_+>1$) and can be computed numerically with very good approximation as the $\blambda_n$ are bounded. 
As a last remark notice that in the uniform case the expression reduces to  \[
\mathscr{G}=\lim_{N\to\infty}\sum_{n=0}^N\frac{\log\lambda_n}{d^n},
\]
where $d$ is the degree of the map and $\lambda_n$s are just numbers. 

\end{document}